\documentclass[12pt, a4paper,chicago]{article}
\usepackage{natbib}
\usepackage{float}
\usepackage{multirow}
\usepackage{amssymb,amsmath, amsfonts}
\usepackage{lscape}
\usepackage{enumerate}
\usepackage{amsthm}
\usepackage[T1]{fontenc}
\usepackage{hyperref}
\usepackage[utf8]{inputenc}
\usepackage{lmodern}
\usepackage[english]{babel}
\usepackage{pdflscape}
\usepackage{csquotes}
\usepackage{mathtools}

\usepackage{xcolor}
\numberwithin{equation}{section}
\usepackage{graphicx}
\usepackage{caption}
\usepackage{subcaption}
\usepackage{upgreek}
\newtheorem{theorem}{Theorem}[section]
\newtheorem{example}{Example}[section]

\newtheorem{lemma}{Lemma}[section]

\setcitestyle{authoryear,open={(},close={)}} 

\begin{document}

\begin{center}
{\Large\bf Dependence and Uncertainty: Information Measures using Tsallis Entropy} \\
\vspace{0.3in}

{ \bf Swaroop Georgy Zachariah$^{a}$,\bf Mohd. Arshad$^{a},$\footnote{Corresponding author. E-mail
addresses: ~arshad@iiti.ac.in (M. Arshad), ~ashokiitb09@gmail.com (Ashok Kumar Pathak), ~swaroopgeorgy@gmail.com
(Swaroop Georgy Zachariah).} and \bf Ashok Kumar Pathak$^{b}$}
 \\
$^a$ Department of Mathematics, Indian Institute of
Technology Indore, India \\
 $^b$ Department of Mathematics and Statistics,
Central University of Punjab, Bathinda,
India
\end{center}

\noindent------------------------------------------------------------------------------------------------------------------------------
\begin{abstract}
In multivariate analysis, uncertainty arises from two sources: the marginal distributions of the variables and their dependence structure. Quantifying the dependence structure is crucial, as it provides valuable insights into the relationships among components of a random vector. Copula functions effectively capture this dependence structure independent of marginals, making copula-based information measures highly significant. However, existing copula-based information measures, such as entropy, divergence, and mutual information, rely on copula densities, which may not exist in many scenarios, limiting their applicability. Recently, to address this issue, \cite{arshad2024multivariateinformationmeasurescopulabased} introduced cumulative copula-based measures using Shannon entropy. In this paper, we extend this framework by using Tsallis entropy, a non-additive entropy that provides greater flexibility for quantifying uncertainties. We propose cumulative copula Tsallis entropy, derive its properties and bounds, and illustrate its utility through examples. We further develop a non-parametric version of the measure and validate it using coupled periodic and chaotic maps. Additionally, we extend Kerridge’s inaccuracy measure and Kullback-Leibler (KL) divergence to the cumulative copula framework. Using the relationship between KL divergence and mutual information, we propose a new cumulative mutual information (CMI) measure, which outperform the limitations of density-based mutual information. Furthermore, we introduce a test procedure for testing the mutual independence among random variables using CMI measure. Finally, we illustrate the potential of the proposed CMI measure as an economic indicator through real bivariate financial time series data.
\end{abstract}
\vskip 2mm


\noindent------------------------------------------------------------------------------------------------------------------------------

\section{Introduction} 
The notion of entropy was introduced by \cite{clausius1850ueber} in relation to the second law of thermodynamics. Later, \cite{boltzmann1872weitere} provided a statistical definition of entropy by linking it to statistical mechanics. \cite{shannon1948mathematical} laid the mathematical foundation of entropy by connecting it with communication theory and it is now popularly known as Shannon entropy. In a probabilistic sense, Shannon entropy is to quantify the uncertainty based on the discrete random variable. Let $X$ be a discrete random variable with mass function $p_i=P(X=x_i),i=1,2,\dots,n$. Then the Shannon entropy is given by
$$\mathcal{H}(X)=-\sum_{i=1}^{n}p_i\log{p_i}.$$
Shannon entropy has various applications in machine learning, reliability theory, physics, chemistry, finance and complex systems.
The continuous counterpart of the Shannon entropy is called differential entropy (DE), which is defined for the absolutely continuous
random variable having probability density function (PDF) $f(\cdot)$ as
\begin{equation}\label{de}
    \mathcal{D}(X)=-\int_{-\infty}^{\infty}f(x)\log{f(x)}dx.
\end{equation}
\noindent However, \cite{rao2004cumulative} pointed out certain limitations of DE and proposed an alternative measure called cumulative residual entropy (CRE). Let $\bar{F}(x)$ be the survival function of a non-negative random variable $X$. Then   CRE can be defined as $$\mathcal{CR}(X)=-\int_{0}^{\infty}\Bar{F}(x)\log{\Bar{F}(x)}dx.$$ 
In a similar line, \cite{di2009cumulative} also proposed the cumulative entropy (CE) by substituting the PDF \( f(x) \) in Eq. (\ref{de}) with the cumulative distribution function (CDF) $F(x)$ of \( X \), which measures the uncertainty in the system's inactivity time. 

\par In the context of thermodynamics, when a system is out of equilibrium or its component states exhibit strong interdependence, non-additive entropy provides a more appropriate measure for quantifying the uncertainty involved in the system. \cite{tsallis1988possible} proposed a non-additive entropy and is defined for an absolutely continuous random variable $X$ with PDF $f(\cdot)$ as
$$\mathcal{T}_{\alpha}(X)=-\int_{-\infty} ^{\infty}f(x)\log_{[\alpha]}(f(x))\, dx
, \quad \alpha\in\mathcal{A},
$$
where $\mathcal{A}=(0,1)\cup (1,\infty)$ and $\log_{[\alpha]}(r)=\frac{r^{\alpha-1}-1}{\alpha-1}$ for every $\alpha\in\mathcal{A}$. It is to be noted that $\underset{\alpha\rightarrow 1}{\lim} \ \log_{[\alpha]}(r)=\log(x)$. Consequently, \( \log_{[\alpha]}(\cdot) \) can be interpreted as a fractional generalization of the standard natural logarithm function. As a result, Tsallis entropy reduces to Shannon entropy when \( \alpha \rightarrow 1 \). Recently, \cite{rajesh2019some} generalized the CRE and proposed cumulative residual Tsallis entropy (CRTE), which is given by
$$\mathcal{TR}_{\alpha}(X)=-\int_{0} ^{\infty}\bar{F}(x)\log_{[\alpha]}(\bar{F}(x))\, dx
, \quad \alpha\in\mathcal{A}.
$$
Similarly, \cite{cali2017some} proposed the cumulative Tsallis entropy (CTE), which generalizes the CE introduced by \cite{di2009cumulative}. Various applications of Tsallis entropy and its variants have been discussed in the literature. For more details, we recommend readers to refer to \cite{cartwright2014roll}, \cite{de2004image}, \cite{sparavigna2015role}, \cite{singh2017tsallis}, \cite{mohamed2022cumulative}, \cite{toomaj2022some}, and the references therein.  Apart from Tsallis entropy, various generalizations of Shannon entropy have been proposed in the literature. For more details, we refer to \cite{renyi1961measures},  \cite{di2007weighted}, \cite{mathai2007pathway}, \cite{psarrakos2017generalized}, \cite{ubriaco2009entropies}, \cite{xiong2019fractional}, and \cite{kayid2022some}.
\par \cite{kullback1951information} introduced a new information measure that is useful in quantifying the divergence between two random variables. This measure is widely known as Kullback-Leibler (KL) divergence and is sometimes referred to as relative entropy. Let \( X_1 \) and \( X_2 \) be two continuous random variables with PDFs \( f_1(x) \) and \( f_2(x) \), respectively. The KL divergence between \( X_1 \) and \( X_2 \) is defined as
\begin{equation}\label{klr}
    KL(f_1 \| f_2) = \int_{-\infty}^{\infty} f_1(x) \log\left(\dfrac{f_1(x)}{f_2(x)}\right) \, dx.
\end{equation}
Note that minimizing the KL divergence between the assumed distribution and empirical distribution is equivalent to maximizing the likelihood of the sample (see \cite[p. 208]{murphy2022probabilistic}). Motivated by the works of \cite{rao2004cumulative}, using survival functions of non-negative random variables, \cite{baratpour2012testing} proposed the cumulative residual KL divergence given by
$$CRKL(\bar{F}_1 \| \bar{F}_2) = \int_{0}^{\infty} \bar{F}_1(x) \log\left(\dfrac{\bar{F}_1(x)}{\bar{F}_2(x)}\right) \, dx-E(X_1)+E(X_2).$$
Moreover, \cite{baratpour2012testing} also discusses the application of the cumulative residual KL (CRKL) divergence for the goodness of fit test for the exponential population. Similarly,
\cite{park2014cumulative} also discussed another alternative of KL divergence using CDF of the random variables. Recently, \cite{mehrali2021parameter} discusses the application of cumulative KL divergence in estimation problems. Furthermore, \cite{mao2020fractional} extended the CRKL divergence using the Tsallis entropy and discusses the application in the finance sector.
\par Apart from entropy, there are several information measures in information theory. One of the popular measures is the inaccuracy measure proposed by \cite{kerridge1961inaccuracy}. Note that in Eq. (\ref{klr}), the KL divergence between two continuous random variables can be written as  $$KL(f_1||f_2)=-\mathcal{D}(X_1)+IN(f_1|f_2),$$
where $IN(f_1|f_2)=-\int_{-\infty}^{\infty}f_1(x)\log(f_2(x))\ dx$ is called the inaccuracy measure suggested by \cite{kerridge1961inaccuracy}. The inaccuracy measure can be interpreted as follows. Let $f_1(x)$ be the true PDF of the data. Suppose, due to experimental error, the experimenter assumes $f_2(x)$ as the PDF of the data instead of $f_1(x)$, then the average uncertainty involved in the incorrect assumption by the experimenter is quantified by the inaccuracy measure $IN(X_1||X_2)$. The cumulative version of the inaccuracy measure is proposed by \cite{kumar2015dynamic}. Recently, \cite{raju2024results} generalizes the cumulative inaccuracy using Tsallis entropy.

\par In literature, the multivariate extension of the existing univariate information measures is also discussed.  For more details we refer to \cite{nadarajah2005expressions},  \cite{ebrahimi2007multivariate}, \cite{rajesh2009bivariate}, \cite{rajesh2014bivariate} and \cite{c2017bivariate}. Recently, the copula function has gained significant popularity in constructing multivariate distributions and modelling the dependencies among random variables. The copula function is the uniform probability measure defined on \( \mathbb{I}^d = [0,1]^d \). \cite{sklar1959fonctions} showed that every joint CDF of a multivariate random variable can be expressed as the function of its marginal CDFs through a copula. 
Furthermore, every joint PDF corresponding to a multivariate random variable can be decomposed into its marginal PDFs and a dependency function that is independent of the marginals, commonly known as the copula density. 
If the copula is absolutely continuous, then the copula density is given by 
\[
c(\mathbf{u}) = \frac{\partial^d C(\mathbf{u})}{\partial u_1 \partial u_2\dots \partial u_d},
\]
where \( \mathbf{u}=(u_1,u_2,\dots,u_d) \in \mathbb{I}^d \).
 For further details, we refer the reader to \cite{nelsen2007introduction}, \cite{trivedi2007copula}, \cite{durante2016principles}, \cite{hofert2018elements}, as well as the recent works of \cite{chesneau2022note}, \cite{Ali08072024}, \cite{zachariah2024new}, and \cite{zachariah2024new1}.
\par In the present paper, we propose a dependence entropy based on copula using Tsallis entropy. A natural question arises regarding the significance of copula-based dependence entropy. It is worth noting that copula-based entropy measures the uncertainty involved in the dependence structure among random variables. In multivariate data analysis, uncertainty associated with a multivariate random variable can be decomposed into two components: the uncertainty due to each marginal distribution and the uncertainty that arises from the dependence structure among the random variables. Note that the copula captures the dependence structure, making copula-based information measures relevant. The scope of copula-based information measures in multivariate data analysis was first discussed by \cite{ma2011mutual}. They showed that the mutual information (MI) of a multivariate random variable is equivalent to the negative of the copula entropy, which is defined as 
\begin{equation}\label{dc}
  \zeta\left(c\right) = -\int_{\mathbb{I}^d} c(\mathbf{u})\log c(\mathbf{u}) \, d\mathbf{u},
\end{equation}
where \( c(\mathbf{u}) \) is the copula density. The mutual information (MI) is one of the important information measures in the multivariate information theory and has numerous practical applications (see \cite{battiti1994using}, \cite{russakoff2004image}, and \cite{kiriakidou2024mutual}). Using the results of \cite{ma2011mutual}, the MI of a multivariate random variable $\bf{X}$ is independent of marginal distributions and depends only on the dependence structure, which is measured by the underlying copula density. Copula entropy has widespread applications across various fields, including science, engineering, hydrology, and finance (see \cite{ce1}, \cite{ce2}, \cite{ce3}).  
However, when the underlying copula is not absolutely continuous, the copula density does not exist, making the copula entropy proposed by \cite{ma2011mutual} inapplicable. Additionally, the copula entropy \( \zeta(c) \) is always negative. Motivated by the works of \cite{rao2004cumulative} and \cite{di2009cumulative}, \cite{sunoj2023survival} replaced the copula density with the  copula function and proposed the cumulative copula entropy (CCE). The bivariate CCE is further extended to higher dimensions by \cite{arshad2024multivariateinformationmeasurescopulabased}. The \( d \)-dimensional cumulative copula entropy is defined as
\[
\xi\left(C\right) = -\int_{\mathbb{I}^d} C(\mathbf{u}) \log C(\mathbf{u}) \, d\mathbf{u},
\]
where \( C(\mathbf{u}) \) represents the copula function. \cite{arshad2024multivariateinformationmeasurescopulabased} also proposed a non-parametric estimator of the CCE using the empirical beta copula and established its almost sure convergence. Additionally, the authors introduced a cumulative copula-based divergence measure derived from Kullback-Leibler (KL) divergence and discussed its applications in copula selection problems and goodness-of-fit tests for copulas. The copula-based inaccuracy measure was first proposed by \cite{hosseini2019results}. Let \( C_1 \) and \( C_2 \) be two \( d \)-dimensional copulas. The copula-based inaccuracy measure is defined as
\[
\mathcal{I}(C_1 \mid C_2) = -\int_{\mathbb{I}^d } C_1(\mathbf{u}) \log\left(C_2(\mathbf{u})\right) \, d\mathbf{u}.
\]
The results were further extended to co-copulas, and the dual of a copula in \cite{hosseini2021discussion}.

\par The existing literature on copula-based information measures remains limited. Motivated by the applicability of non-additive Tsallis entropy, this paper seeks to quantify the uncertainty associated with the dependence structure of multivariate random variables through Tsallis entropy. Furthermore, the paper emphasizes the importance of copula-based information measures by illustrating their relevance and utility in practical applications. The main contributions of this paper are as follows:

\begin{itemize}
    \item We propose copula-based information measures using Tsallis entropy and refer to the proposed dependence entropy as cumulative copula Tsallis entropy (CCTE).
    \item The proposed non-additive entropy generalizes the results of \cite{arshad2024multivariateinformationmeasurescopulabased} and validates its applicability in the context of Rulkov maps within chaos and bifurcation theory.
    \item We introduce a non-parametric estimator for the CCTE and establish its almost sure convergence.
    \item A new inaccuracy measure for copulas is proposed, along with an exploration of its mathematical properties. This inaccuracy measure extends the work presented in \cite{hosseini2021discussion}.
    \item Inspired by \cite{mao2020fractional}, we propose a cumulative copula Tsallis divergence derived from cumulative Tsallis divergence.
    \item To address cases where the copula density may not exist, we introduce a new mutual information measure, called cumulative mutual information measure, that does not rely on copula density, using the relationship between KL divergence and mutual information.
    \item Two specific applications of the proposed mutual information measure are discussed:
    \begin{enumerate}
        \item Testing the independence of several random variables.
        \item Analyzing multivariate financial time series, where the proposed MI serves as an economic indicator.
    \end{enumerate}
\end{itemize}

\par The remaining structure of this paper is organized into three main parts. In Section \ref{sec1}, we introduce the cumulative copula Tsallis entropy, examine its mathematical properties, and provide examples using well-known copulas. Section \ref{sec2} presents a non-parametric estimator for the proposed dependence entropy, provides a theoretical proof of its
almost sure convergence, and validates the results using Monte Carlo simulations. Section \ref{sec3} discusses the validation of the proposed dependence entropy by applying it to Rulkov maps.
The second part of the paper, starting with Section \ref{sec4}, introduces a copula-based inaccuracy measure and explores related inequalities and ordering properties. Section \ref{sec5} presents a newly developed cumulative copula divergence based on the Tsallis divergence, highlighting its properties and introducing a new mutual information measure.
The final part of the paper discusses the applications of the proposed mutual information measure. Section \ref{sec6} is divided into two subsections. Subsection \ref{subsec1} proposes a new testing procedure for the mutual independence among the components of a multivariate random variable. The proposed test is compared with existing procedures based on Cram\'{e}r-von Mises and Kolmogorov-Smirnov distance measures. The proposed test is applied to real data to demonstrate its practical utility. Subsection \ref{subsec2} illustrates the use of the proposed mutual information measure as an economic indicator in analyzing multivariate financial time series.
The paper concludes in Section \ref{sec7} with a summary of the findings and a discussion of potential directions for future research.
\section{Cumulative Copula Tsallis Entropy}\label{sec1}
In this section, we propose the cumulative copula Tsallis entropy (CCTE), defined as follows

    \begin{equation}
        \xi_{\alpha}(C) =-\int_{\mathbb{I}^d} C(\mathbf{u})\log_{[\alpha]}(C(\mathbf{u}))  \, d\mathbf{u}, \quad \alpha\in\mathcal{A},
        \label{tcce1}
    \end{equation}
where $\log_{[\alpha]}(r)=\frac{r^{\alpha-1}-1}{\alpha-1}$ and $\mathcal{A}=(0,1)\cup (1,\infty)$. It is easy to show that for any $\alpha\in\mathcal{A}$, the function $h(r)=-r\log_{[\alpha]}(r)$ is bounded by $0\leq h(r)\leq \alpha^{\alpha/1-\alpha}$ for every $u\in\mathbb{I}$. It follows that  $0\leq \xi_{\alpha}(C)\leq \alpha^{1/1-\alpha}\leq 1$, for every $\alpha\in\mathcal{A}.$
Moreover, 
$$\underset{\alpha \rightarrow 1}{\lim} \ \xi_{\alpha}(C) =  -\int_{\mathbb{I}^d} C(\mathbf{u}) \log \left(C(\mathbf{u})\right) d\mathbf{u}=\xi(C).$$

In the following subsection, we present typical examples of the CCTE for various well-known bivariate and multivariate copulas.
\subsection{Examples}
\begin{example}
For any bivariate copula, the Fr{\'e}chet-Hoeffding lower bound copula, defined as 
\[
W(u_1, u_2) = \max\{u_1 + u_2 - 1, 0\},
\]
provides the lower bound for every bivariate copula. The CCTE corresponding to the Fr{\'e}chet-Hoeffding lower bound copula is given by
\begin{align*}
\xi_{\alpha}(W) 
&= -\int_0^1 \int_0^1 \max\{u_1 + u_2 - 1, 0\} 
\log_{[\alpha]}\big(\max\{u_1 + u_2 - 1, 0\}\big) \, du_1 \, du_2 \\
&= \frac{1}{\alpha - 1} \int_{0}^1 \int_{0}^{u_1} 
\big(u_2 - u_2^{\alpha}\big) \, du_2 \, du_1 \\
&= \frac{\alpha + 4}{6(\alpha + 1)(\alpha + 2)}.
\end{align*}
\end{example}

\begin{example}
 The CCTE for the Marshall-Olkin copula is given by
\begin{align*}
\xi_{\alpha}(W)
&= \frac{1}{\alpha - 1} \bigg[
\int_{0}^1 \int_0^{u_2^{q/p}} 
\big( u_1 u_2^{1-q} - \left(u_1 u_2^{1-q}\right)^{\alpha} \big) \, du_1 \, du_2 \int_{0}^1 \int_0^{u_1^{p/q}} 
\big( u_1^{1-p} u_2 - \left(u_1^{1-p} u_2\right)^{\alpha} \big) \, du_2 \, du_1 
\bigg] \\
&= \frac{\left(p + q\right) \big(\omega(1) - \omega(\alpha)\big)}{\alpha^2 - 1},
\end{align*}
where $\omega(x) = \frac{1}{(x+1)(p+q) - x p q}$.
\end{example}


\begin{example}
The underlying copula corresponding to the mutual independence of random variables is the product copula, defined as 
\[
\Pi(\mathbf{u}) = u_1 u_2 \ldots u_d.
\]
The CCTE for the product copula is given by
\begin{align*}
\xi_{\alpha}(\Pi) 
&= \frac{1}{\alpha - 1} \int_0^1 \int_0^1 \cdots \int_0^1 \big(u_1 u_2 \cdots u_d - (u_1 u_2 \cdots u_d)^{\alpha}\big) \, du_1 \, du_2 \ldots du_d \\  
&= \frac{(\alpha + 1)^d - 2^d}{2^d (\alpha^2 - 1) (\alpha + 1)^{d-1}}.
\end{align*}
\end{example}

\begin{example}
    For any $d$-dimensional copula, the Fr{\'e}chet-Hoeffding upper bound copula, defined as 
\[
M({\bf{u}})=\min\{u_1,u_2,\dots, u_d\},
\]
provides the upper bound for every $d$-dimensional copula. The CCTE corresponding to the Fr{\'e}chet-Hoeffding upper bound copula is given by
\begin{align*}
\xi_{\alpha}(M) 
&= \frac{1}{\alpha - 1} \int_{0}^{1} \int_{0}^{1} \cdots \int_{0}^{1} 
\min\{u_1, u_2, \dots, u_k\} - \left(\min\{u_1, u_2, \dots, u_d\}\right)^{\alpha} \, du_1 \, du_2 \ldots du_d \\
&= \frac{d}{\alpha - 1} \int_{0}^{1} \big(u - u^{\alpha}\big)(1 - u)^{d-1} \, du \\
&= \frac{d}{\alpha - 1} \left(\beta(2, d) - \beta(\alpha + 1, d)\right),
\end{align*}
where $\beta(p, q)$ is the well-known beta function. Note that the transformation of the above multiple integrals into a single integral uses the concept of order statistics. The multiple integral in the above equation can be expressed as 
$E\big(U_{[1]} - U_{[1]}^{\alpha}\big),$
where \( U_1, U_2, \dots, U_k \) are \( d \) random samples from the uniform distribution over \(\mathbb{I}\), and \( U_{[1]} = \min\{U_1, U_2, \dots, U_d\} \). 
\end{example}
\noindent Next, we explore several inequalities associated with the CCTE, which establishes the bounds for the measure.
\subsection{Inequalities}
\begin{theorem}\label{th2.1}
For every \(d\)-dimensional copula \(C\) with \(\xi_{\alpha}(C)\), the following inequalities hold:
\[
\xi_\alpha(C) \begin{cases} 
 \geq \xi(C), & \text{if } \alpha \in (0,1), \\
 \leq \xi(C), & \text{if } \alpha \in (1,\infty).
\end{cases}
\]
\end{theorem}

\begin{proof}
For any \(r \geq 0\), it holds that \(1 - r \leq -\log(r)\). Consequently, for any \(r \in \mathbb{I}\),
\[
-r\log_{[\alpha]}(r) = \frac{r\left(1 - r^{\alpha - 1}\right)}{\alpha - 1} \begin{cases} 
 \geq -r\log(r), & \text{if } \alpha \in (0,1), \\
 \leq -r\log(r), & \text{if } \alpha \in (1,\infty).
\end{cases}
\]
The result follows by substituting \(r = C(\mathbf{u})\) and integrating over \(\mathbb{I}^d\).
\end{proof}

\begin{theorem}\label{th2.2}
Let \(\xi_{\alpha}(C)\) be the CCTE of a copula \(C\), then
\[
\xi_\alpha(C) \begin{cases} 
 \geq \xi_2(C), & \text{if } \alpha \in (0,2]\setminus\{1\}, \\
 \leq \xi_2(C), & \text{if } \alpha \in (2,\infty).
\end{cases}
\]
\end{theorem}

\begin{proof}
For \(\alpha \in (0,2]\setminus\{1\}\), the function 
\[
g(r) = -\log_{[\alpha]}(r) + \log_{[2]}(r) = \frac{1 - r^{\alpha - 1}}{\alpha - 1} - 1 + r
\]
attains its minimum at \(r = 1\). For \(\alpha \in (2,\infty)\), \(g(r)\) attains its maximum at \(r = 1\). Thus, \(g(r) \geq 0\) if \(\alpha \in (0,2]\setminus\{1\}\) and \(g(r) \leq 0\) if \(\alpha \in (2,\infty)\). Substituting \(r = C(\mathbf{u})\) and integrating over \(\mathbb{I}^d\), the result follows.
\end{proof}

Let $C(\mathbf{u})$ be a $d$-dimensional copula. The multivariate version of Spearman's correlation can be defined as 
\begin{align}
   \rho_d(C)=&c_d\left(2^d\int_{\mathbb{I}^d}C(\mathbf{u})d\mathbf{u}-1\right),\label{spm1}
\end{align}
where $c_d=\dfrac{d+1}{2^d-d-1}$. For more details, we refer to \cite{schmid2010copula} and \cite{bedHo2016multivariate}).
The following theorem provides the relation between multivariate Spearman's correlation coefficient and CCTE. 

\begin{theorem}
Let $C$ be a $d$-dimensional copula with multivariate Spearman's correlation coefficient    $\rho_d(C)$. Then for any $\alpha\in\mathcal{A},$
    $$
   \xi_{\alpha}(C)\leq g_d(C)\log_{[\alpha]}\left(g_d(C)\right),
    $$ 
   where $ g_d(C)=\left(\rho_d(C)+c_d\right)c_d^{-1}2^{-d} .$
\end{theorem}
\begin{proof}
For any $\alpha\in\mathcal{A}$, $h(r)=-r\log_{[\alpha]}(r)=\dfrac{r-r^{\alpha}}{\alpha-1}$ is concave for $r\in\mathbb{I}$. The result follows, using Jensen's inequality on the concave function.
\end{proof}
Now, we will focus on the CCTE of the weighted arithmetic mean (W.A.M.) of copulas. It is important to note that the W.A.M of copulas with the same dimension is also a copula. The following theorem shows the uncertainty involved in the W.A.M. of copulas.
\begin{theorem}
Let \( C_1, C_2, \dots, C_p \) be \( p \) copulas of dimension \( d \) with corresponding CCTE values \( \xi_{\alpha}(C_1), \xi_{\alpha}(C_2), \dots, \xi_{\alpha}(C_p) \). Define \( C^{\Sigma}(\mathbf{u}) = \sum_{j=1}^p l_j C_j(\mathbf{u}) \) as the W.A.M. of these copulas, where \( l_j \in \mathbb{I} \) for \( j = 1, 2, \dots, p \) and \( \sum_{j=1}^p l_j = 1 \). Let \( \xi_{\alpha}(C) \) denote the CCTE of \( C \). Then the following inequality holds
\[
\sum_{j=1}^p l_j \, \xi_{\alpha}(C_j) \leq \xi_{\alpha}(C^{\Sigma}).
\]
\end{theorem}

\begin{proof}
The function \( h(r) = -r\log_{[\alpha]}(r) \) is concave, which implies that 
\[
\sum_{j=1}^p l_j h(r_j) \leq h\left( \sum_{j=1}^p l_j r_j \right),
\]
for every $r_j\in\mathbb{I}$. The result follows by substituting \( r_j = C_j(\mathbf{u}) \) and integrating over \( \mathbb{I}^d \).

\end{proof}

\begin{figure}[ht]
    \centering
    \includegraphics[width=0.5\linewidth]{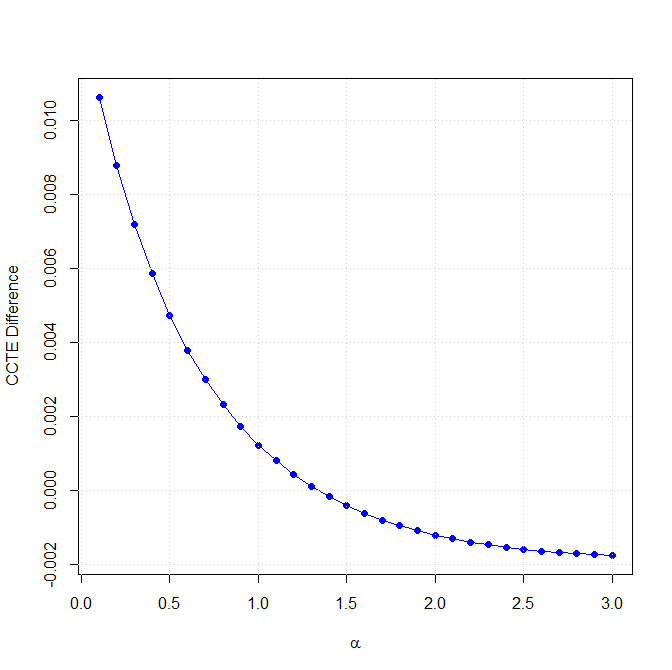}
    \caption{\( \xi_{\alpha}(C_1) - \xi_{\alpha}(C_2) \) for different values of $\alpha$}
    \label{ccted1}
\end{figure}

   Let \( C_1(\mathbf{u}) \) and \( C_2(\mathbf{u}) \) be two \( d \)-dimensional copulas. Then, \( C_1 \) is less positive lower orthant dependent (PLOD) than \( C_2 \), denoted by \( C_1 \overset{\text{PLOD}}{\prec} C_2 \), if \( C_1(\mathbf{u}) \leq C_2(\mathbf{u}) \) for every \( \mathbf{u} \in \mathbb{I}^d \). Now, we will show that PLOD ordering does not necessarily imply the corresponding CCTE ordering through a counterexample.  For a counterexample, take
\[
C_1(u_1, u_2) = \left( 1 + \left[ (u_1^{-1} - 1)^2 + (u_2^{-1} - 1)^2 \right]^{0.5} \right)^{-1},
\]
and 
\[
C_2(u_1, u_2) = \min\{ u_1, u_2 \}.
\]
The difference \( \xi_{\alpha}(C_1) - \xi_{\alpha}(C_2) \) is shown in Figure \ref{ccted1}, which illustrates that the inequality is not preserved for PLOD ordering.

\noindent In the following subsection, we establish the uniform convergence property of CCTE.
\subsection{Uniform Convergence}
\begin{theorem}
    Let $\{C_N\}$ be a sequence of $d$-dimensional copulas with CCTE $\xi_{\alpha}(C_N)$, and let \( C \) be a $d$-dimensional copula with CCTE $\xi_{\alpha}(C)$. If \( C_N \) converges uniformly to \( C \), then $\xi_{\alpha}(C_N)$ converges uniformly to $\xi_{\alpha}(C)$ for all $\alpha\in\mathcal{A}$.
\end{theorem}

\begin{proof}
The function \( h(r) = -r\log_{[\alpha]}(r)\) is bounded and uniformly continuous on \( \mathbb{I} \). Thus, for any \( \delta > 0 \), there exists \( \eta > 0 \) such that for any \( r_1, r_2 \in \mathbb{I} \) satisfying \( |r_1 - r_2| < \eta \), we have
\begin{equation}\label{unifc1}
    \left|h(r_1) - h(r_2)\right| < \delta.
\end{equation}
Substituting \( r_1 = C_N(\mathbf{u}) \) and \( r_2 = C(\mathbf{u}) \) in Eq.~(\ref{unifc1}), it follows that
\begin{equation}\label{cnu}
    \left|h\left(C_N(\mathbf{u})\right) - h\left(C(\mathbf{u})\right)\right| < \delta,
\end{equation}
whenever
\[
    \left|C_N(\mathbf{u}) - C(\mathbf{u})\right| < \eta.
\]
If \( C_N \) converges uniformly to \( C \), then for any \( \eta > 0 \), there exists a natural number \( m \geq N \) such that 
\begin{equation}\label{unifc}
    \left|C_N(\mathbf{u}) - C(\mathbf{u})\right| < \eta,
\end{equation}
for every \( \mathbf{u} \in \mathbb{I}^d \). Using Eq.~(\ref{cnu}) and Eq.~(\ref{unifc}), and applying the bounded convergence theorem, the result follows.
\end{proof}
\section{Empirical Cumulative Copula Tsallis Entropy}\label{sec2}
In this section, we use the empirical copula to propose a non-parametric estimator for CCTE. A non-parametric estimate of CCE based on the empirical copula was introduced by \cite{sunoj2023survival}.  Let ${\bf{X}_j}=\left(X_{j1}, X_{j2},\dots,X_{jd}\right);j=1,2,\dots,n$ be a random sample of size $n$ from a multivariate population.  Based on these samples, the empirical copula \( \hat{C}_n \) can be used to estimate the underlying copula, defined as 
\begin{equation}\label{empirical}
    \hat{C}_n\left(\mathbf{u}\right) =\dfrac{1}{n}\sum_{j=1}^n\prod_{k=1}^{d}\mathbf{I}\left(\dfrac{R_{jk}}{n+1}\leq u_{k}\right),
\end{equation}
where \( R_{jk} \) is the rank of the \( k \)-th component of the  \( j \)-th observation \( X_{jk} \), and \( \mathbf{I}(\cdot) \) denotes the indicator function (see \cite{deheuvels1979fonction}, \cite{nelsen2007introduction}, \cite{GF2}, and \cite{durante2016principles}). Now, using the definition of empirical copula, we define the empirical CCTE as
\begin{equation}\label{eccte}
    \xi_{\alpha}(\hat{C}_n)=-\int_{\mathbb{I}^d}\hat{C}_n(\mathbf{u})\log_{[\alpha]}\left(\hat{C}_n(\mathbf{u})\right)\,d\mathbf{u}, \quad \alpha\in\mathcal{A}.
\end{equation}
The following theorem provides the upper bound for the empirical CCTE.
\begin{theorem}
    Let \( \hat{C}_n \) be the empirical copula based on the random sample ${\bf{X}_1},{\bf{X}_2}\dots,{\bf{X}_n}$  from a multivariate distribution of dimension \( d \). Let \( \xi_{\alpha}(\hat{C}_n) \) be the empirical CCTE defined in Eq.~(\ref{eccte}). Then, for any \( \alpha \in \mathcal{A} \),
    \[
    \xi_{\alpha}(\hat{C}_n) \leq -\frac{1}{n}\mathcal{R}\log_{[\alpha]}\left(\mathcal{R}\right), 
    \]
    where $\mathcal{R}=\left\{\displaystyle\dfrac{1}{n} \sum_{j=1}^n \prod_{k=1}^d \left(1 - \frac{R_{jk}}{n+1}\right)\right\}.$
\end{theorem}
\begin{proof}
    By Jensen's inequality, we have
    \begin{align*}
        \xi_{\alpha}(\hat{C}_n) &\leq \frac{1}{\alpha - 1} \left\{ \int_{\mathbb{I}^d} \hat{C}_n(\mathbf{u}) \, d\mathbf{u} - \left( \int_{\mathbb{I}^d} \hat{C}_n(\mathbf{u}) \, d\mathbf{u} \right)^{\alpha} \right\} \\
        &= \frac{1}{\alpha - 1} \left\{ \int_{\mathbb{I}^d} \frac{1}{n} \sum_{j=1}^n \prod_{k=1}^d \mathbf{I} \left( \frac{R_{jk}}{n+1} \leq u_k \right) d\mathbf{u} - \left( \int_{\mathbb{I}^d} \frac{1}{n} \sum_{j=1}^n \prod_{k=1}^d \mathbf{I} \left( \frac{R_{jk}}{n+1} \leq u_k \right) d\mathbf{u} \right)^{\alpha} \right\} \\
        &= \frac{1}{\alpha - 1} \left\{ \frac{1}{n} \sum_{j=1}^n \prod_{k=1}^d \left( 1 - \frac{R_{jk}}{n+1} \right) - \left( \frac{1}{n} \sum_{j=1}^n \prod_{k=1}^d \left( 1 - \frac{R_{jk}}{n+1} \right) \right)^{\alpha} \right\}\\
        &=-\frac{1}{n}\mathcal{R}\log_{[\alpha]}\left(\mathcal{R}\right).
    \end{align*}
\end{proof}
We now focus on the consistency of the proposed non-parametric estimator. The following theorem asserts the convergence of the empirical CCTE.
\begin{theorem}
  The empirical CCTE converges to the true CCTE almost surely. Specifically, for any \( \alpha \in \mathcal{A} \),
  as \( n \rightarrow \infty \), we have
  \[
  \hat{\xi}_{\alpha}(C_n) \rightarrow \xi_{\alpha}(C) \quad \text{a.s}.
  \]
\end{theorem}
\begin{proof}
By the Glivenko-Cantelli theorem for empirical copulas, we have that as \( n \rightarrow \infty \),
\begin{equation}\label{s1}
\sup_{\mathbf{u} \in \mathbb{I}^d} \left| C_n(\mathbf{u}) - C(\mathbf{u}) \right| \rightarrow 0, \quad \text{a.s.}
\end{equation}
For further details, see \cite{deheuvels1979fonction}, \cite{ec1}, \cite{ec2}, and \cite{ec3}. 
Using the continuous mapping theorem of almost sure convergence, along with the bounded convergence theorem, the result follows.
\end{proof}

Now, we illustrate this theorem through a simulation study for various copulas, specifically considering the following:

- Clayton copula:
  \[
  C(\mathbf{u}) = \max\left\{\sum_{i=1}^d u_i^{\theta} - d + 1, 0\right\}^{-1/\theta}, \quad \theta \in [-1, \infty) \setminus \{0\}.
  \]

- Gumbel-Hougaard copula:
  \[
  C(\mathbf{u}) = \exp\left\{-\left(\sum_{i=1}^d (-\log(u_i))^{\theta}\right)^{1/\theta}\right\}, \quad \theta \geq 1.
  \]

\begin{figure}[ht]
    \centering
    \subfloat[Clayton copula with parameter $\theta=1.5$]{\includegraphics[height=5cm, width=5cm]{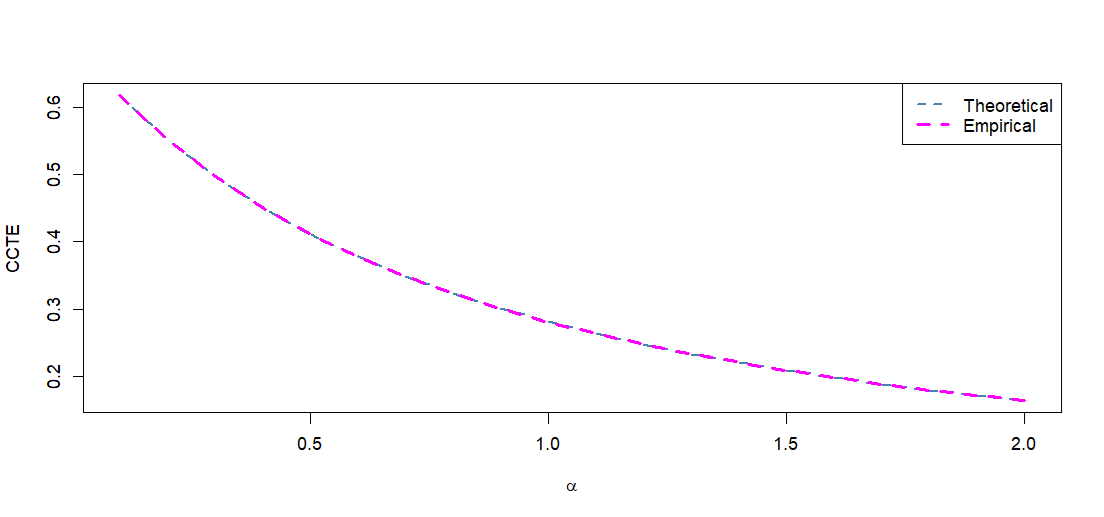}} \hfill
    \subfloat[Gumbel-Hougaard copula with parameter $\theta=2$]{\includegraphics[height=5cm, width=5cm]{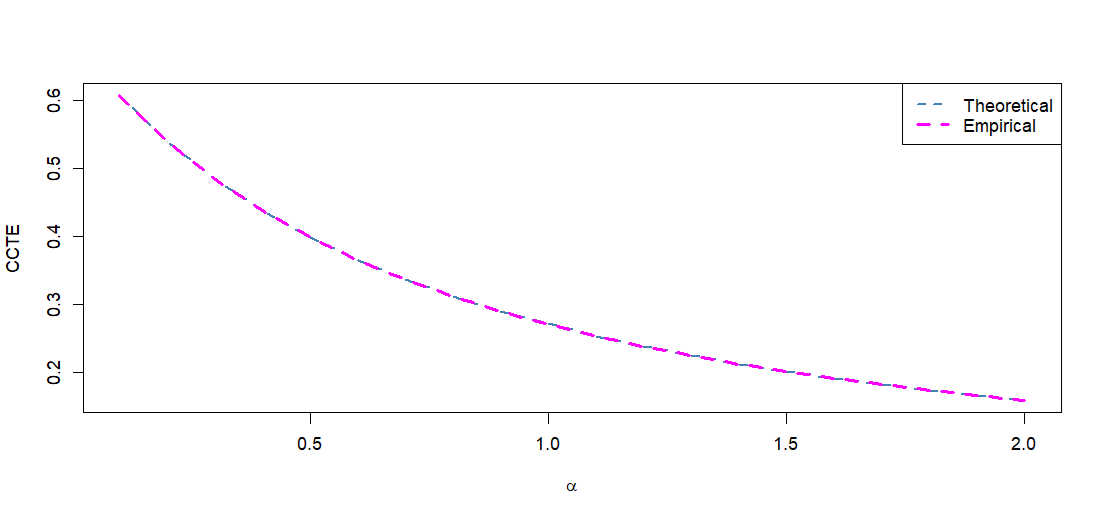}} \hfill
    \subfloat[Frank copula with parameters $\theta=2.5$]{\includegraphics[height=5cm, width=5cm]{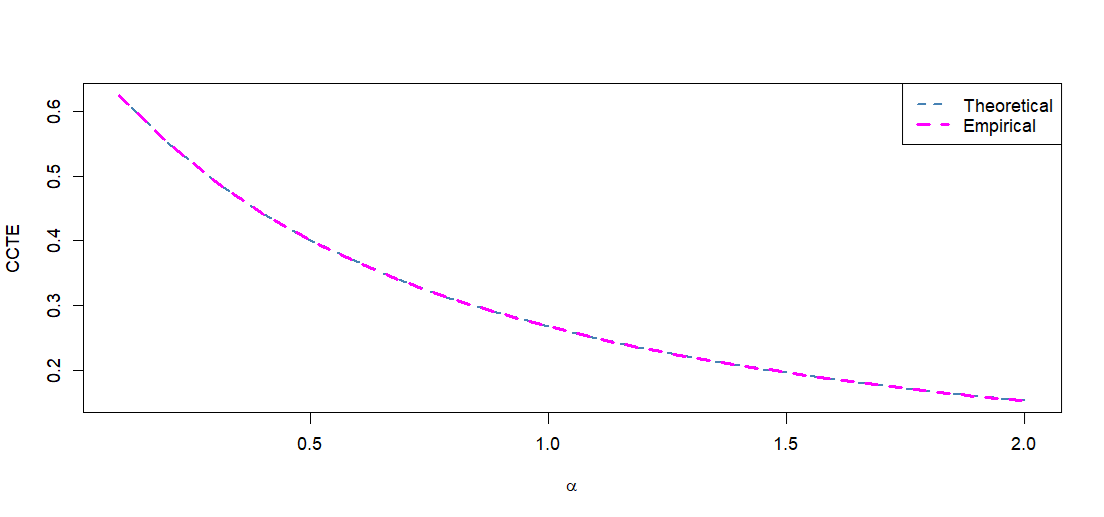}}
    \caption{The empirical CCTE and theoretical CCE of various bivariate copulas.}\label{fig1}
\end{figure}

\begin{figure}[ht]
    \centering
    \subfloat[Clayton copula with parameter $\theta=1.5$]{\includegraphics[height=5cm, width=5cm]{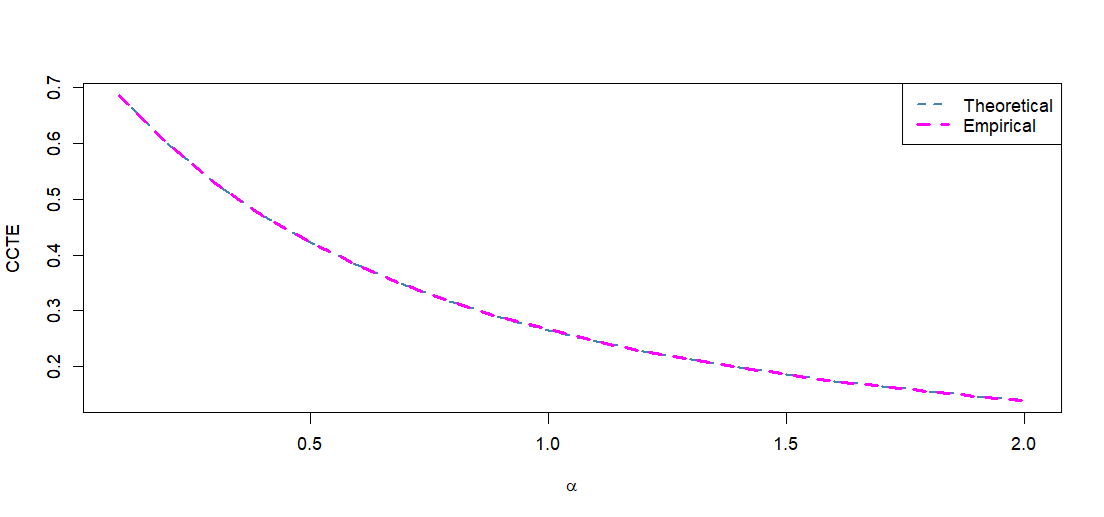}} \hfill
    \subfloat[Gumbel-Hougaard copula with parameter $\theta=2$]{\includegraphics[height=5cm, width=5cm]{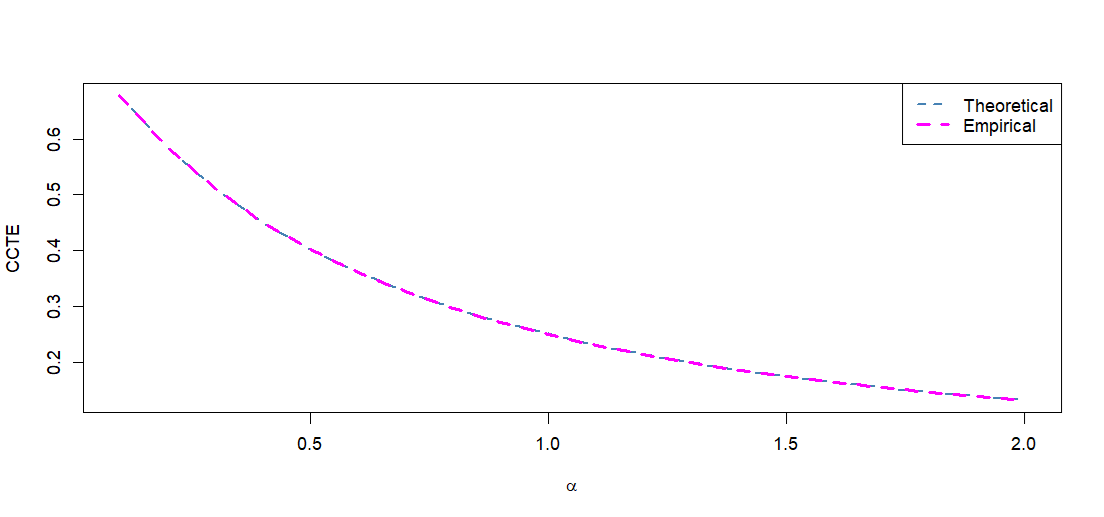}} \hfill
    \subfloat[Frank copula with parameters $\theta=2.5$]{\includegraphics[height=5cm, width=5cm]{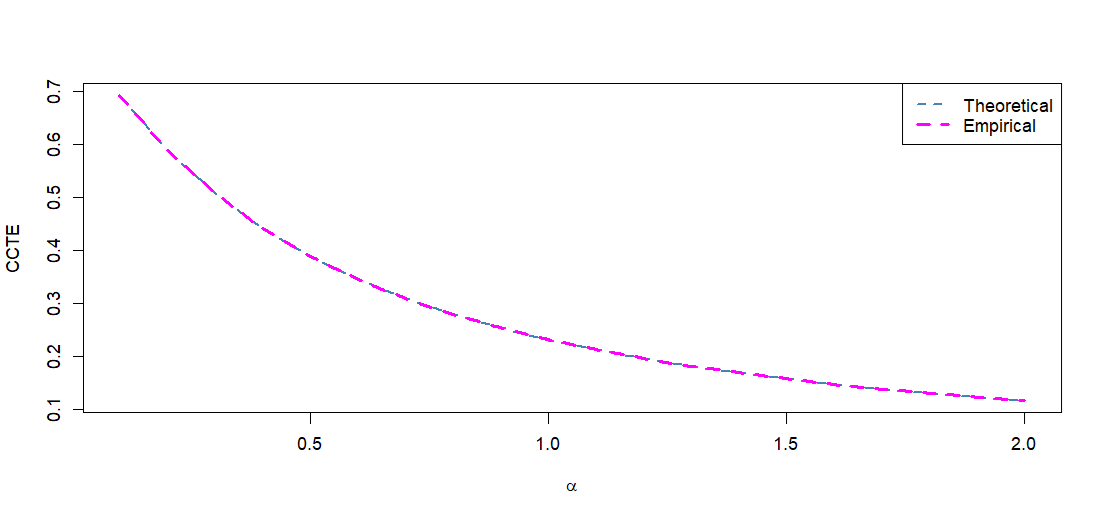}}
    \caption{The empirical CCTE and theoretical CCE of various trivariate copulas.}\label{fig2}
\end{figure}

- Frank copula:
  \[
  C(\mathbf{u}) = -\frac{1}{\theta} \log\left(1 + \frac{\prod_{i=1}^d e^{-\theta u_i} - 1}{e^{-\theta} - 1}\right), \quad \theta \in \mathbb{R} \setminus \{0\}.
  \]
We generated 1,000 random numbers from each of the copulas mentioned above and computed the empirical CCTE, comparing these estimates with the actual values. Due to the absence of a closed-form expression for the empirical CCTE, we evaluated the integrals numerically using the \texttt{adaptIntegrate} function from the \texttt{cubature} package in R (version 4.2.2). Figures \ref{fig1} and \ref{fig2} illustrate the convergence of the non-parametric estimate of CCTE for the Clayton copula, Gumbel-Hougaard copula, and Frank copula in both bivariate and trivariate cases. From these figures, it is evident that the shape of the CCTE varies with the dimension of the copula.

\section{Validity of Cumulative Copula Tsallis Entropy with Chaotic Theory}\label{sec3}
Here, we validate our entropy measure using chaotic theory. We consider the identical Rulkov maps given by the system of equations:
\begin{align*}
x_{n+1} &= \frac{\delta}{1+x_n^2} + \beta + \gamma(y_n - x_n), \\
y_{n+1} &= \frac{\delta}{1+y_n^2} + \beta + \gamma(x_n - y_n),
\end{align*}
where \( \gamma \) is the coupling parameter. For more details, refer to \cite{rulkov2001regularization} and \cite{r1}. It has been shown that for specific values of \( \delta = 2, 0.2, -0.8, -2, -2.5 \), the coupled map exhibits periodicity of \( 1 \), \( 2 \), \( 4 \), and two chaotic sequences, respectively. 

\begin{figure}[ht]
    \centering
    \includegraphics[width=1\linewidth]{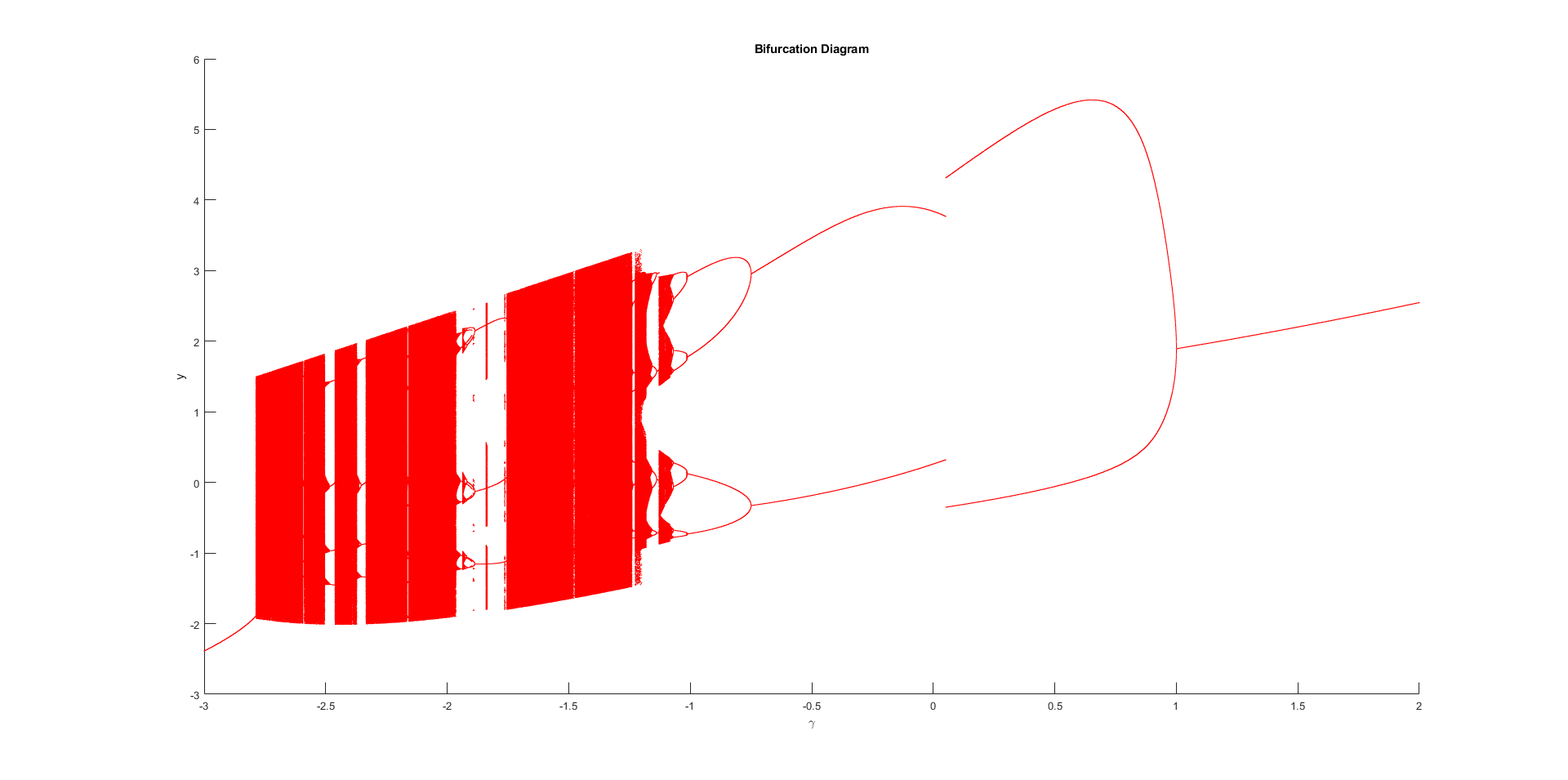}
    \caption{Bifurcation diagram of identical Rulkov maps}
    \label{Bif}
\end{figure}

\begin{figure}[ht]
    \centering
    \includegraphics[width=1\linewidth]{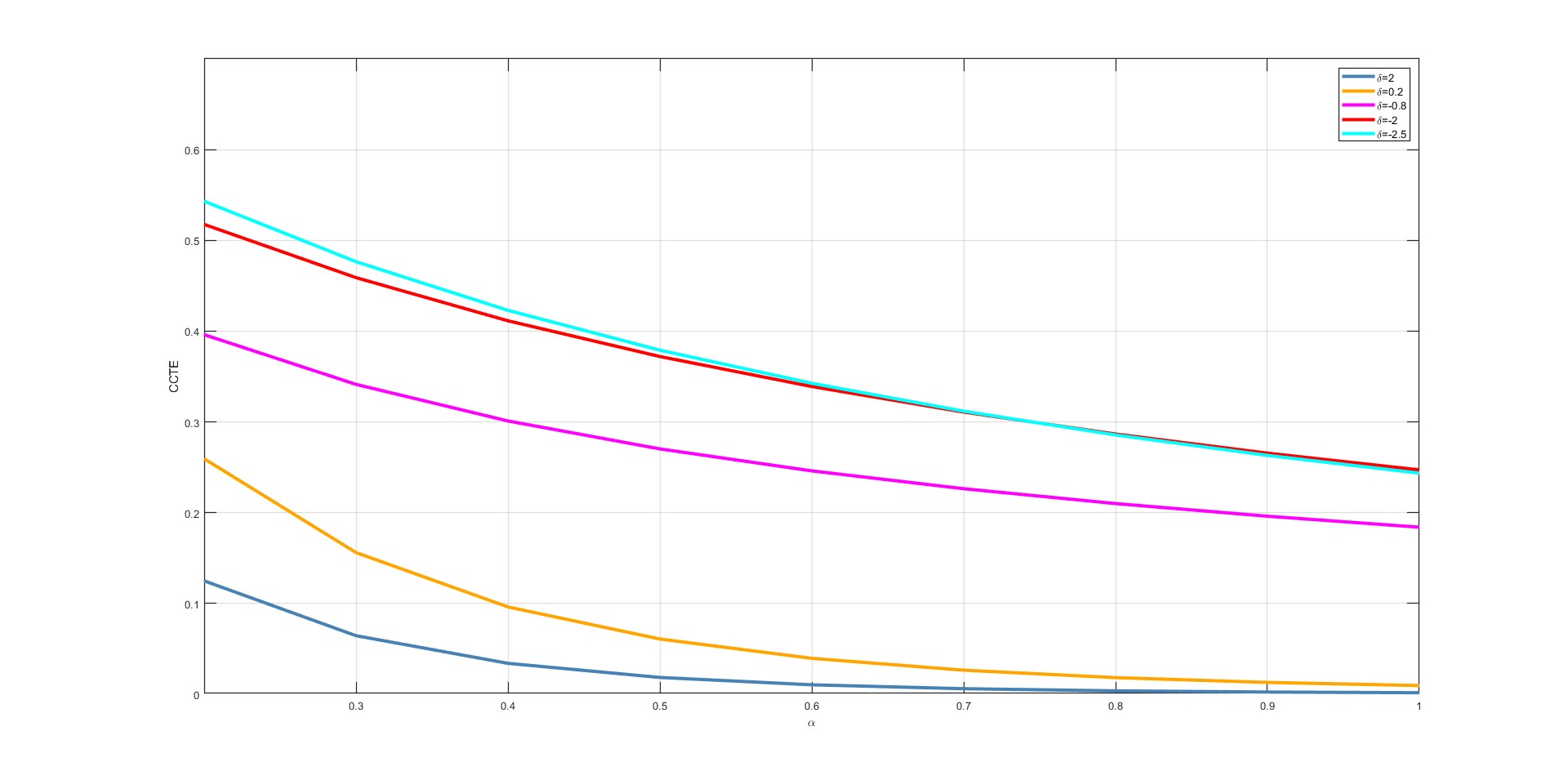}
    \caption{CCTE of identical Rulkov maps}
    \label{chaos}
\end{figure}

We perform numerical simulations on the Rulkov maps and consider the first 2000 observations with initial values \( x_0 = 0.1 \) and \( y_0 = 0.5 \). We plot the bifurcation diagram, which is presented in Figure \ref{Bif}, for verification purposes. Using the empirical copula, we calculate the empirical CCTE. As per the theory, for periodic cases, the dependence entropy tends to be lower. Even as the period increases, the CCTE increases, while in chaotic cases, the CCTE remains higher than in the periodic cases. The results are shown in Figure \ref{chaos}, where we observe that the CCTE increases with the periodicity and is greater in the chaotic case than in the periodic case.

 \section{Cumulative Copula Tsallis Inaccuracy Measure}\label{sec4}
Apart from entropy, several information measures for uncertainty are available in the literature. In this section, we introduce a new measure known as the cumulative copula Tsallis inaccuracy (CCTI) measure, which generalizes the inaccuracy measure proposed by \cite{hosseini2019results}. Let \( C_1(\mathbf{u}) \) and \( C_2(\mathbf{u}) \) be two copulas of the same dimension \( d \). If an experimenter uses \( C_2 \) to model the dependence structure among random variables instead of the true copula \( C_1 \), the copula-based inaccuracy measure quantifies the error introduced by this incorrect assumption is well-known as misspecification in literature. This incorrect assumption may be due to experimental error or wrong observations, or maybe both. Let \( C_1 \) be the true copula, and suppose the experimenter uses \( C_2 \) instead of \( C_1 \). Then, the CCTI measure corresponds to the copula $C_1$ and $C_2$ is defined as 
\begin{equation}\label{ccti}
    \mathcal{I}_{\alpha}(C_1|C_2) = -\int_{\mathbb{I}^d}C_1(\mathbf{u})\log_{[\alpha]} \left(C_2(\mathbf{u})\right)\, d\mathbf{u}.
\end{equation}
Note that $\underset{\alpha\rightarrow1}{\lim } \ \mathcal{I}_{\alpha}(C_1|C_2)=\mathcal{I}(C_1|C_2)$ (copula-based inaccuracy measure proposed by \cite{hosseini2019results}), and when $C_1=C_2=C$ then $\mathcal{I}_{\alpha}(C_1|C_2)=\xi_{\alpha}(C)$. Thus, the proposed inaccuracy measure can be viewed as a generalization of the inaccuracy measure proposed by \cite{hosseini2019results}, and the parameter $\alpha$ will give flexibility in quantifying the experimental error. We will now go through a few examples based on commonly used copula in the literature.
\begin{example}
    The CCTI measure corresponding to the Fr{\'e}chet-Hoeffding lower bound copula \( W(u_1, u_2) \) and the product copula \( \Pi(u_1, u_2) \) is given by
    \[
        \mathcal{I}_{\alpha}(W|\Pi)
        = \frac{1}{6(\alpha-1)} + \frac{\beta(\alpha,\alpha+2) + (\alpha+1)\beta(\alpha,2) - 1}{\alpha(\alpha^2-1)}.
    \]
    The Fr{\'e}chet-Hoeffding lower bound copula is used for modelling strongly negatively dependent bivariate data. Thus, the above inaccuracy measure quantifies the uncertainty involved in incorrectly assuming independence when the data exhibits strong negative dependence.
\end{example}

\begin{example}
    Consider the FGM copula given by:
    \[
        C(u_1,u_2) = u_1u_2\left(1 + \theta (1-u_1)(1-u_2)\right),
    \]
    where \( \theta \in [-1,1] \). The CCTI measure corresponding to the FGM copula and the product copula is given by
    \[
        \mathcal{I}_{\alpha}(C|\Pi) = \frac{\theta+9}{36(\alpha-1)} - \frac{1}{(\alpha^2-1)(\alpha+1)} - \frac{\theta \, \beta(\alpha+1,2)}{\alpha-1}.
    \]
\end{example}

\begin{example}
    The CCTI measure corresponding to the Fr{\'e}chet-Hoeffding upper bound copula \( M(u_1, u_2) \) and the product copula is given by
    \[
        \mathcal{I}_{\alpha}(M|\Pi) = \frac{1}{(d+1)(\alpha-1)} + \frac{d!}{(\alpha^2-1)\prod_{j=2}^d(j\alpha+1)}.
    \]
    The Fr{\'e}chet-Hoeffding upper bound copula is used for modelling strongly positively dependent data. The above inaccuracy measure quantifies the uncertainty involved in incorrectly assuming independence when the data exhibits strong positive dependence.
\end{example}

\begin{example}
    The \( d \)-variate Cuadras-Aug{\'e} copula, proposed by \cite{cuadras2009constructing}, is given by
    \begin{equation}\label{cu}
        C(\mathbf{u}) = \prod_{i=1}^d u_{[i]}^{\gamma_i},
    \end{equation}
    where \( \gamma_1, \gamma_2, \dots, \gamma_d \) are copula parameters such that \( C(\mathbf{u}) \) in Eq.~\eqref{cu} is a valid copula, and \( u_{[1]}, u_{[2]}, \dots, u_{[d]} \) are the ordered values of \( u_1, u_2, \dots, u_d \) in ascending order. For more details, see \cite{nadarajah2017compendium} and \cite{cuadras2009constructing}. The CCTI measure corresponding to the Cuadras-Aug{\'e} copula and the product copula is given by
    \[
        \mathcal{I}_{\alpha}(\Pi|C) = \frac{1}{2^d(\alpha-1)} - \frac{1}{\alpha-1}\prod_{j=1}^d \frac{1}{\left(\sum_{i=1}^j \delta(i) + j\right)},
    \]
    where $\delta(i)$ satisfies the recurrence relation $\delta(i)=\delta(i-1)+(\alpha-1)\gamma_i+2$ for $i=2,\dots,d$ with $\delta(1)=\theta_1(\alpha-1)+2$.
\end{example}

We now discuss some mathematical properties of the CCTI measure. Similar to Theorem \ref{th2.1} and \ref{th2.2}, we have the following result. The proof is similar to the proof of Theorem \ref{th2.1} and \ref{th2.2}, so we omitted.
\begin{theorem}
Let $C_1$ and $C_2$ be two copulas of the same dimension. Let $\mathcal{I}_{\alpha}(C_1|C_2)$ be the inaccuracy measure by incorrect use of $C_2$, instead of $C_1$. Then
$$ \mathcal{I}_{\alpha}(C_1|C_2)\begin{cases} 
 \geq \mathcal{I}_{}(C_1|C_2), & \text{if } \alpha \in (0,1), \\
 \leq \mathcal{I}(C_1|C_2), & \text{if } \alpha \in (1,\infty).
\end{cases}$$
\end{theorem}

\begin{theorem}
 Let $\mathcal{I}_{\alpha}(C_1|C_2)$ be the  CCTI measure with respect to the copulas $C_1$ and $C_2$ of the same dimension $d$, then the following inequalities hold.
$$ \mathcal{I}_{\alpha}(C_1|C_2)\begin{cases} 
 \geq \mathcal{I}_{2}(C_1|C_2), & \text{if } \alpha \in (0,2]\setminus\{1\}, \\
 \leq \mathcal{I}_{2}(C_1|C_2), & \text{if } \alpha \in (2,\infty).
\end{cases}$$
\end{theorem}

\begin{theorem}
Let \( C_1, C_2, \dots, C_p \) be \( p \) \( d \)-dimensional copulas, and let \( C^{\Sigma}(\mathbf{u}) = \sum_{j=1}^p l_j C_j(\mathbf{u}) \) be the W.A.M. of these copulas, where \( l_j \in \mathbb{I} \) for \( j = 1, 2, \dots, p \) with \( \sum_{j=1}^p l_j = 1 \). Let \( C \) be any \( d \)-dimensional copula, then 
\[
\mathcal{I}_{\alpha}\left(C;C^{\Sigma}\right) \begin{cases} 
 \leq \sum_{j=1}^{p} l_j \mathcal{I}_{\alpha}(C;C_j), & \text{if } \alpha \in (0,2]\setminus\{1\}, \\
 \geq \sum_{j=1}^{p} l_j \mathcal{I}_{\alpha}(C;C_j), & \text{if } \alpha \in (2,\infty).
\end{cases}
\]
\end{theorem}

\begin{proof}
Since the function \( -\log_{[\alpha]}(y) = \frac{1 - y^{\alpha - 1}}{\alpha - 1} \) is convex (concave) in \( y \geq 0\) if \( \alpha \in (0,2]\setminus\{1\} \) \( (\alpha \in (2,\infty)) \), it follows that for fixed \( x \in \mathbb{I} \),
\[
 -x\log_{[\alpha]}\left(z\right) \begin{cases} 
 \leq \sum_{j=1}^{p} l_j\left(  -x\log_{[\alpha]}(y_j)\right), & \text{if } \alpha \in (0,2]\setminus\{1\}, \\
 \geq \sum_{j=1}^{p} l_j \left( -x\log_{[\alpha]}(y_j)\right), & \text{if } \alpha \in (2,\infty),
\end{cases}
\]
where $z=\sum_{j=1}^p l_j y_j$  and \( y_1,y_2,\dots,y_p \in \mathbb{I} \). Substituting \( x = C(\mathbf{u}) \) and \( y_j = C_j(\mathbf{u}) \) for every \( j = 1,2,\dots,p, \) and integrating over \( \mathbb{I}^d \), we obtain the required result.
\end{proof}

Now, we will discuss the inaccuracy measure related to the weighted geometric mean (W.G.M.) of copulas. Let \( C_1, C_2, \dots, C_p \) represent \( p \) \( d \)-dimensional copulas. The W.G.M. of these copulas is defined as:
\begin{equation}\label{wgm}
    C^{\Pi}(\mathbf{u}) = \prod_{j=1}^p C_j(\mathbf{u})^{q_j},
\end{equation}
where \( q_j \in \mathbb{I} \) for \( j = 1, 2, \dots, p \), and \( \sum_{j=1}^p q_j = 1 \). It is important to note that \( C^{\Pi}(\mathbf{u}) \) defined in Eq.~\eqref{wgm} is not always a valid copula. However, under specific conditions, it can satisfy the requirements of a copula. For more details, see \cite{cuadras2009constructing}, \cite{zhang2013some}, \cite{diaz2022extension}, and \cite{zachariah2024new1}. 

The following theorem provides an upper bound for the inaccuracy measure associated with the W.G.M. of copulas.

\begin{theorem}
    Let \( C^{\Pi}(\mathbf{u}) = \prod_{j=1}^p C_j(\mathbf{u})^{q_j} \) be the W.G.M. of \( p \) copulas of dimension \( d \) defined in Eq.~\eqref{wgm}, and let \( C \) be any \( d \)-dimensional copula. Then
    \[
        \mathcal{I}_{\alpha}\left(C^{\Pi}; C\right) \leq \prod_{j=1}^p\left[\mathcal{I}_{\alpha}\left(C_j; C\right)\right]^{q_j}.
    \]
\end{theorem}

\begin{proof}
    Let \( g : \mathbb{I}^d \to \mathbb{R}^{+} \) be a function. For any \( t \neq 0 \), we have
    \begin{align}
        \int_{\mathbb{I}^d} g(\mathbf{u}) \left[C^{\Pi}(\mathbf{u})\right]^t d\mathbf{u}
        &= \int_{\mathbb{I}^d} \prod_{j=1}^p g(\mathbf{u})^{q_j} \left[C_j(\mathbf{u})^{q_j t}\right] d\mathbf{u}. \nonumber
    \end{align}
    By applying the generalized H\"older's inequality (see \cite{kufner1977function}, \cite{finner1992generalization}), we obtain
    \begin{align}
        \int_{\mathbb{I}^d} g(\mathbf{u}) \left[C^{\Pi}(\mathbf{u})\right]^t d\mathbf{u}
        &\leq \prod_{j=1}^p \left( \int_{\mathbb{I}^d} g(\mathbf{u}) \left[C_j(\mathbf{u})\right]^t d\mathbf{u} \right)^{q_j}. \label{win}
    \end{align}
The CCTI measure associated with \( C^{\Pi} \) and \( C \) is given by
    \begin{align*}
        \mathcal{I}_{\alpha}\left(C^{\Pi}; C\right) 
        &= \int_{\mathbb{I}^d}C^{\Pi}(\mathbf{u}) \log_{[\alpha]}\left(C(\mathbf{u}\right) \, d\mathbf{u}.
    \end{align*}
Substituting \( g(\mathbf{u}) = \log_{[\alpha]}\left(C(\mathbf{u}\right)\) and \( t = 1 \) into inequality \eqref{win}, we obtain
    \begin{align*}
        \mathcal{I}_{\alpha}\left(C^{\Pi}; C\right)
        &\leq \prod_{j=1}^p \left( \int_{\mathbb{I}^d} \frac{1 - C^{\alpha-1}(\mathbf{u})}{\alpha-1} \cdot C_j(\mathbf{u}) d\mathbf{u} \right)^{q_j} \\
        &= \prod_{j=1}^p \left(\mathcal{I}_{\alpha}\left(C_j; C\right)\right)^{q_j}.
    \end{align*}
This completes the proof.
\end{proof}

Now, we will discuss some results for CCTI based on the PLOD property of copulas. 
\begin{theorem}
    Let \( C_1 \overset{\text{PLOD}}{\prec} C_2 \). Then, for any \( \alpha \in \mathcal{A} \), the inequality
    $\mathcal{I}_{\alpha}(C_1|C_2) \leq \mathcal{I}_{\alpha}(C_2| C_1)$
    holds.
\end{theorem}
\begin{proof}
  By the assumption \( C_1 \overset{\text{PLOD}}{\prec} C_2 \), for any \( \alpha \in \mathcal{A} \), we have  
\begin{align*}
    \mathcal{I}_{\alpha}(C_1| C_2) - \mathcal{I}_{\alpha}(C_2| C_1) 
    &= \int_{\mathbb{I}^d} \frac{C_1(\mathbf{u}) \left(1 - C_2^{\alpha-1}(\mathbf{u})\right)}{\alpha-1} 
    - \frac{C_2(\mathbf{u}) \left(1 - C_1^{\alpha-1}(\mathbf{u})\right)}{\alpha-1} \, d\mathbf{u} \\
    &\leq \int_{\mathbb{I}^d} C_1(\mathbf{u}) 
    \left(\frac{C_1^{\alpha-1}(\mathbf{u}) - C_2^{\alpha-1}(\mathbf{u})}{\alpha-1}\right) \, d\mathbf{u} \\
    &\leq 0.
\end{align*}

\end{proof}

\begin{theorem}\label{th4.6}
   Let $C_1$ and $C_2$ and $C_3$ be three $d$-dimensional copulas. If \( C_1 \overset{\text{PLOD}}{\prec} C_2 \), then for any $\alpha\in\mathcal{A}$, the following triangle inequalities hold.
   \begin{enumerate}[(a)]
       \item $\mathcal{I}_{\alpha}(C_3| C_1)+\mathcal{I}_{\alpha}(C_1| C_2)\geq \mathcal{I}_{\alpha}(C_3| C_2)$
       \item $\mathcal{I}_{\alpha}(C_1| C_2)+\mathcal{I}_{\alpha}(C_2| C_3)\geq \mathcal{I}_{\alpha}(C_1| C_3)$.
   \end{enumerate}
\end{theorem}
\begin{proof}
   We will prove the part (a) of the theorem. The proof of the part (b) is similar to that of the first part and is therefore omitted. Under the assumption of \( C_1 \overset{\text{PLOD}}{\prec} C_2 \), we have
   \begin{align*}
      \mathcal{I}_{\alpha}(C_3| C_1)+\mathcal{I}_{\alpha}(C_1| C_2)=& \int_{\mathbb{I}^d} C_1(\mathbf{u})\log_{[\alpha]}\left(C_2(\mathbf{u})\right)-C_3(\mathbf{u})\log_{[\alpha]}\left(C_1(\mathbf{u})\right) \, d\mathbf{u}\\=&\int_{\mathbb{I}^d} \frac{C_3(\mathbf{u}) \left(1 - C_1^{\alpha-1}(\mathbf{u})\right)}{\alpha-1} 
    - \frac{C_1(\mathbf{u}) \left(1 - C_2^{\alpha-1}(\mathbf{u})\right)}{\alpha-1} \, d\mathbf{u}\\
    \geq& \int_{\mathbb{I}^d} \left(C_3(\mathbf{u})+C_1(\mathbf{u}\right)\left(\frac{1-C_2^{\alpha-1}(\mathbf{u})}{\alpha-1}\right)\, d\mathbf{u}\\
    \geq&\int_{\mathbb{I}^d}C_3(\mathbf{u})\left(\frac{1-C_2^{\alpha-1}(\mathbf{u})}{\alpha-1}\right)\, d\mathbf{u}= \mathcal{I}_{\alpha}(C_3| C_2).
   \end{align*}

\end{proof}
\noindent The proofs of the following theorem is similar to the proof of Theorem \ref{th4.6}, so we left out here.
\begin{theorem}
    Let \( C_1 \), \( C_2 \), and \( C_3 \) be three copulas of same dimension. Then for any $\alpha\in\mathcal{A}$, we have
    \begin{enumerate}[(a)]
        \item  If \( C_1 \overset{\text{PLOD}}{\prec} C_2 \), then $\mathcal{I}_{\alpha}(C_1| C_3) \leq \mathcal{I}_{\alpha}(C_2| C_3)$ and $\mathcal{I}_{\alpha}(C_3| C_2) \leq \mathcal{I}_{\alpha}(C_3| C_1)$.
    \item If \( C_1\overset{\text{PLOD}}{\prec} C_2 \) and \( C_1\overset{\text{PLOD}}{\prec} C_3 \), then
    $\mathcal{I}_{\alpha}(C_1| C_3) \leq \mathcal{I}_{\alpha}(C_2| C_3)\leq \mathcal{I}_{\alpha}(C_2| C_1) $.
    \item If \( C_1\overset{\text{PLOD}}{\prec} C_3 \) and \( C_2\overset{\text{PLOD}}{\prec} C_3 \), then
    $\mathcal{I}_{\alpha}(C_1| C_3) \leq \mathcal{I}_{\alpha}(C_1| C_2)\leq \mathcal{I}_{\alpha}(C_3| C_1) $.
    \item If \( C_1\overset{\text{PLOD}}{\prec}C_2\overset{\text{PLOD}}{\prec} C_3 \), then 
    $$\max\left\{I_{\alpha}(C_2|C_1),I_{\alpha}(C_3|C_2)\right\}\leq \mathcal{I}_{\alpha}(C_3|C_2)\quad 
    \text{and}\quad
     \min\left\{I_{\alpha}(C_1|C_2),I_{\alpha}(C_2|C_3)\right\}\geq \mathcal{I}_{\alpha}(C_1|C_3).$$
    \end{enumerate}

\end{theorem}
\section{Cumulative Copula Tsallis  Divergence and Mutual Information} \label{sec5}
In this section, we propose a new divergence measure between two copulas based on Tsallis divergence, along with a new mutual information (MI) measure derived from the cumulative copula. The concept of divergence plays a crucial role in the field of statistics, particularly in statistical inference. 

Let \( \mathbf{X_1} \) and \( \mathbf{X_2} \) be two multivariate random variables with identical marginals but with underlying copulas that are not necessarily the same. Let \( \mathbf{f_1}(\cdot) \) and \( \mathbf{f_2}(\cdot) \) denote the joint PDF of \( \mathbf{X_1} \) and \( \mathbf{X_2} \) and, \( c_1 \) and \( c_2 \) denote the underlying copula densities corresponding to \( \mathbf{X_1} \) and \( \mathbf{X_2} \), respectively. Then, the KL divergence between \( \mathbf{X_1} \) and \( \mathbf{X_2} \) is equivalent to the KL divergence between the two copula densities, as discussed in \cite{ghosh2024copula}. That is,
\[
KL(\mathbf{f_1}||\mathbf{f_2}) = \int_{\mathbb{I}^d} c_1(\mathbf{u}) \log\left(\frac{c_1(\mathbf{u})}{c_2(\mathbf{u})}\right) d\mathbf{u}.
\]
As outlined in the introduction, the above copula density divergence may not be suitable in certain cases. \cite{arshad2024multivariateinformationmeasurescopulabased} introduced an alternative divergence measure between copulas based on the KL divergence, called cumulative copula Kullback-Leibler divergence (CCKLD). Let \( C_1 \) and \( C_2 \) be two copulas of the same dimension. The CCKLD between \( C_1 \) and \( C_2 \) is defined as
\begin{equation*}
\Delta(C_1||C_2) = \int_{\mathbb{I}^d} C_1(\mathbf{u}) \log\left(\frac{C_1(\mathbf{u})}{C_2(\mathbf{u})}\right) d\mathbf{u} 
- \left[\frac{\rho_d(C_1) - \rho_d(C_2)}{2^d c_d}\right],
\end{equation*}
where \( \rho_d(\cdot) \) is the multivariate Spearman's correlation defined in Eq. (\ref{spm1}). 
\par Motivated by the works of \cite{mao2020fractional} and \cite{arshad2024multivariateinformationmeasurescopulabased}, we now propose the following divergence measure between two copulas based on Tsallis divergence:
\begin{equation}\label{cctkl}
\Delta_{\alpha}(C_1||C_2) = \int_{\mathbb{I}^d} C_1(\mathbf{u}) \log_{[\alpha]}\left(\frac{C_1(\mathbf{u})}{C_2(\mathbf{u})}\right) d\mathbf{u} 
- \left[\frac{\rho_d(C_1) - \rho_d(C_2)}{2^d c_d}\right],
\end{equation}
where \( \alpha \in \mathcal{A} \). We refer to \( \Delta_{\alpha}(C_1||C_2) \) as the cumulative copula Tsallis divergence (CCTD).
 It is straightforward to show that $$\underset{\alpha\rightarrow 1}{\lim} \ \Delta_{\alpha}(C_1||C_2)=\Delta(C_1||C_2).$$ Moreover, when $\alpha=2$, CCTD reduces to $$\Delta_{2}(C_1||C_2)=\int_{\mathbb{I}^d}\dfrac{\left(C_1(\mathbf{u})-C_2(\mathbf{u})\right)^2}{C_2(\mathbf{u})} \ d\mathbf{u},$$ which we call as the $\chi^2$ divergence between two copulas, $C_1$ and $C_2$. We denote it as $\chi^2(C_1||C_2)$. The $\chi^2$ divergence between two copula densities is discussed in \cite{ghosh2024copula}. As the copula density may not exist in certain cases, the proposed measure can be considered as an alternative. The following theorem shows that CCTD is always non-negative and zero whenever $C_1=C_2$ almost surely. 
 \begin{theorem}
 Let  $\Delta_{\alpha}(C_1||C_2)$ be the CCTD between two copulas $C_1$ and $C_2$, then for any $\alpha\in\mathcal{A}$, $\Delta_{\alpha}(C_1||C_2)$ is always non-negative, and $\Delta_{\alpha}(C_1||C_2)=0$ whenever $C_1=C_2$ almost surely.

 \end{theorem}
 \begin{proof}
 By definition, we have
\begin{align*}
   \Delta_{\alpha}(C_1||C_2)=& \int_{\mathbb{I}^d} C_1(\mathbf{u})\log_{[\alpha]}\left(\frac{C_1(\mathbf{u})}{C_2(\mathbf{u})}\right)d\mathbf{u}-\left[\dfrac{\rho_d(C_1)-\rho_d(C_2)}{2^dc_d}\right]\\
   =& \int_{\mathbb{I}^d} C_1(\mathbf{u})\log_{[\alpha]}\left(\frac{C_1(\mathbf{u})}{C_2(\mathbf{u})}\right)- C_1(\mathbf{u})+C_2(\mathbf{u}) \ d\mathbf{u}\\
=&\int_{\mathbb{I}^d}C_1(\mathbf{u})f_{\alpha}\left(\dfrac{C_2(\mathbf{u})}{C_1(\mathbf{u})}\right) \ d \mathbf{u},
\end{align*} 
where $f_{\alpha}(r)=\dfrac{r^{1-\alpha}-1}{\alpha-1}+r-1$.
Using elementary calculus, one can easily show that for any $\alpha\in\mathcal{A}$, the function $f_{\alpha}(r)$ is always non-negative for every $r\geq 0$ and $f(r)$ attains its minimum at $r=1$. It follows that $C_1(\mathbf{u})f_{\alpha}\left(\dfrac{C_2(\mathbf{u})}{C_1(\mathbf{u})}\right)$ is always non-negative and equal to zero if and only if $C_1(\mathbf{u})=C_2(\mathbf{u})$ for every $u\in\mathbb{I}^d$, which concludes the proof.
 \end{proof}
\noindent Now, we will discuss a few mathematical properties associated with the proposed divergence measure. The following theorem discusses how CCTD relates to CCKLD and $\chi^2$ divergence. 
\begin{theorem}
Let $C_1$ and $C_2$ be two copulas of the same dimension, then the following inequalities hold.
\begin{enumerate}[(a)]
    \item  $\Delta_{\alpha}(C_1||C_2)\begin{cases} 
 \geq \Delta(C_1||C_2), & \text{if } \alpha \in (0,1), \\
 \leq \Delta(C_1||C_2), & \text{if } \alpha \in (1,\infty).
\end{cases}$

\item $ \Delta_{\alpha}(C_1||C_2)\begin{cases} 
 \geq \chi^2(C_1||C_2), & \text{if } \alpha \in (0,2]\setminus\{1\}, \\
 \leq \chi^2(C_1||C_2), & \text{if } \alpha \in (2,\infty).
\end{cases}$
\end{enumerate}
\end{theorem}
\begin{theorem}
Let \( C^{\Sigma}(\mathbf{u}) = \sum_{j=1}^p l_j C_j(\mathbf{u}) \) represent the W.A.M. of \( p \) copulas, \( C_1, C_2, \dots, C_p \), of the dimension $d$, where \( l_j \in \mathbb{I} \) for \( j = 1, 2, \dots, p \), satisfying \( \sum_{j=1}^p l_j = 1 \). Let \( C \) be any \( d \)-dimensional copula, then for any $\alpha\in\mathcal{A}$, we have
\begin{enumerate}[(a)]
    \item $\Delta_{\alpha}\left(C||C^{\Sigma}\right)\leq \sum_{j=1}^{p} l_j \Delta_{\alpha}\left(C||C_j\right)$
    \item $\Delta_{\alpha}\left(C^{\Sigma}||C\right)\leq \sum_{j=1}^{p} l_j \Delta_{\alpha}\left(C_j||C\right).$
\end{enumerate}
\end{theorem}
\begin{proof}[Proof of (a)]  
For any fixed $\alpha\in\mathcal{A}$, the function $f_{\alpha}(r)=\dfrac{r^{1-\alpha}-1}{\alpha-1}+r-1$ is a convex function for every $x\geq0$. It follows that for every $x_j\geq0$, $j=1,2,\dots,p$, we have $f_{\alpha}\left(\sum_{j=1}^pl_jr_j\right)\leq \sum_{j=1}^pf_{\alpha}(l_jr_j)$. Substituting $r_j=\dfrac{C_2(\mathbf{u})}{C_1(\mathbf{u})}$ and using the definition of CCTD, we obtain 
\begin{align*}
   \Delta_{\alpha}\left(C||C^{\Sigma}\right)= &\int_{\mathbb{I}^d}C(\mathbf{u})f_{\alpha}\left(\dfrac{C^{\Sigma}(\mathbf{u})}{C(\mathbf{u})}\right) \ d \mathbf{u}\\
   \leq& \sum_{j=1}^p\int_{\mathbb{I}^d}C(\mathbf{u})f_{\alpha}\left(\dfrac{C_j(\mathbf{u})}{C(\mathbf{u})}\right) \ d \mathbf{u}\\
   =&\sum_{j=1}^{p} l_j \Delta_{\alpha}\left(C||C_j\right).
\end{align*}
\end{proof}  

\begin{proof}[Proof of (b)]  For any $\alpha\in\mathcal{A}$, we define the function $g_{\alpha}(r)=\dfrac{k^{1-\alpha}r^{\alpha}-r}{\alpha-1}+r-k$, where $k\geq0$. It is easy to show that the $g_{\alpha}(r)$ is a convex function for every $x\geq0$. Now, substituting $k=C_2(\mathbf{u})$ and $r=C_1(\mathbf{u})$ and the similar argument of the proof of part (a), we obtain the required result. 
\end{proof}
\noindent Now, we will discuss the ordering property of CCTD based on the PLOD ordering of copula.
\begin{theorem}
    If \( C_1 \overset{\text{PLOD}}{\prec} C_2 \), then
     $\Delta_{\alpha}(C_1||C_2)\begin{cases} 
 \geq \Delta_{\alpha}(C_2||C_1), & \text{if } \alpha \in \left(0,\frac{1}{2},\right] \\
 \leq \Delta_{\alpha}(C_2||C_1), & \text{if } \alpha \in \left(\frac{1}{2},\infty\right)\setminus\{1\}.
\end{cases}$
\end{theorem}
\begin{proof}
For every fixed \(\alpha \in \mathcal{A}\), define the function \(h_\alpha : \mathbb{I} \to \mathbb{R}\) as
\[
h_\alpha(r) = r \log_{[\alpha]}(r) - \log_{[\alpha]}\left(\frac{1}{r}\right) - 2r + 2, \quad r \in \mathbb{I}.
\]
It is easy to show that $h_\alpha(r)$ is an increasing (decreasing) function in $r\in\mathbb{I}$ if $\alpha \in \left(0,\frac{1}{2}\right] \ \left(\alpha \in \left(\frac{1}{2},\infty\right)\setminus\{1\}\right)$. It follows that
 for every $r \in \mathbb{I}$, we have
 \begin{equation}\label{cases}
   h_\alpha(r) \begin{cases} 
\geq 0, &  \text{ if } \alpha \in \left(0, \frac{1}{2}\right], \\
\leq 0, &  \text{ if } \alpha \in \left(\frac{1}{2}, \infty\right)\setminus\{1\}.
\end{cases}  
 \end{equation}
  Note that if \( C_1 \overset{\text{PLOD}}{\prec} C_2 \), then $\Delta_{\alpha}(C_1||C_2) - \Delta_{\alpha}(C_2||C_1)=\int_{\mathbb{I}^d}C_2(\mathbf{u})h_\alpha\left(\frac{C_1(\mathbf{u})}{C_2(\mathbf{u})}\right)\, d\mathbf{u}$. Now, the result follows from inequality (\ref{cases}).
\end{proof}
\noindent Analogous to Theorem \ref{th4.6}, we also have triangle inequality for CCTD. 
\begin{theorem}
   Let $C_1$ and $C_2$ and $C_3$ be three $d$-dimensional copulas. \begin{enumerate}[(a)]
       \item If \( C_1(\mathbf{u})\leq \min\{C_2(\mathbf{u}),C_3(\mathbf{u})\}\) for every $\mathbf{u}\in\mathbb{I}^d$, then $\Delta_{\alpha}(C_3|| C_1)+\Delta_{\alpha}(C_1|| C_2)\geq \Delta_{\alpha}(C_3|| C_2)$
       \item If \( C_1 \overset{\text{PLOD}}{\prec} C_2\overset{\text{PLOD}}{\prec}C_3 \), then 
     $\Delta_{\alpha}(C_1|| C_2)+\Delta_{\alpha}(C_2|| C_3)\leq \Delta_{\alpha}(C_1|| C_3)$.
   \end{enumerate}

\end{theorem}
\begin{proof}
   We will prove the first part of the theorem. Since the second part of the proof is similar to the first part, so we left out here. Assume that \( C_1(\mathbf{u})\leq \min\{C_2(\mathbf{u}),C_3(\mathbf{u})\}\) for every $\mathbf{u}\in\mathbb{I}^d$. Now consider
   \begin{align*}
     \Delta_{\alpha}(C_3|| C_1)+\Delta_{\alpha}(C_1|| C_2)- \Delta_{\alpha}(C_3|| C_2)& =\int_{\mathbb{I}^d}  C_3(\mathbf{u})\log_{[\alpha]}\left(\frac{C_3(\mathbf{u})}{C_1(\mathbf{u})}\right)+ C_1(\mathbf{u})\log_{[\alpha]}\left(\frac{C_1(\mathbf{u})}{C_2(\mathbf{u})}\right)-C_3(\mathbf{u})\log_{[\alpha]}\left(\frac{C_3(\mathbf{u})}{C_2(\mathbf{u})}\right)\, d\mathbf{u}\\
    &= \int_{\mathbb{I}^d} \left(\frac{C_3^{\alpha}(\mathbf{u})-C_1^{\alpha}(\mathbf{u})}{\alpha-1}\right)\left[\dfrac{1}{C_1^{\alpha-1}(\mathbf{u})}-\dfrac{1}{C_2^{\alpha-1}(\mathbf{u})}\right]\, d\mathbf{u}\\
    &\geq0.
   \end{align*}

\end{proof}
Now, we proceed to discuss the mutual information (MI) of a multivariate random vector. Let \( X_1 \) and \( X_2 \) be two continuous random variables with joint PDF \( f(x_1, x_2) \) and marginal PDFs \( f_1(x_1) \) and \( f_2(x_2) \), respectively. The MI between \( X_1 \) and \( X_2 \) is defined as 
\[
MI(f_1; f_2) = \int_{-\infty}^{\infty}\int_{-\infty}^{\infty} f(x_1, x_2) \log\left(\frac{f(x_1, x_2)}{f_1(x_1)f_2(x_2)}\right) \, dx_1 dx_2.
\]
Note that \( MI(f_1; f_2) \) is equivalent to the KL divergence between the joint PDF of \( (X_1, X_2) \) and the product of the marginal PDFs of \( X_1 \) and \( X_2 \). For further details, we refer readers to \cite{cover1999elements}, \cite{ash2012information}, and \cite{murphy2022probabilistic}. 

\cite{joe1987majorization} extended the notion of mutual information to higher dimensions. Let \( \mathbf{X} \) be a \( d \)-variate continuous random variable with joint PDF \( f(\cdot) \) and marginal CDFs (PDFs) \( F_i(\cdot) \) (\( f_i(\cdot) \)), \( i = 1, 2, \dots, d \), where the marginals need not be identical. Let \( c(\cdot) \) denote the copula density corresponding to \( \mathbf{X} \). By Sklar's theorem, the joint PDF \( f(\mathbf{x}) \) can be expressed as 
\[
f(\mathbf{x}) = c\left(F_1(x_1), F_2(x_2), \dots, F_d(x_d)\right) \prod_{j=1}^d f_j(x_j),
\]
where \( \mathbf{x} \in \mathbb{R}^d \). It follows that the MI corresponding to \( \mathbf{X} \) is given by 
\[
MI(f_1; f_2; \dots; f_d) = \int_{\mathbb{I}^d} c(\mathbf{u}) \log\left(c(\mathbf{u})\right) d\mathbf{u}.
\]
The relationship between MI and copula entropy has been discussed independently by \cite{blumentritt2012mutual} and \cite{ma2011mutual}. However, the term ``copula entropy" was first introduced in \cite{ma2011mutual}. 

If the underlying copula is not absolutely continuous (e.g., the minimum copula), the copula density does not exist, and estimating MI non-parametrically in such cases becomes challenging. Using the relationship between KL divergence and MI, and based on the proposed cumulative copula Tsallis divergence (CCTD), we introduce an alternative MI measure called cumulative mutual information (CMI) of order \( \alpha \). Let \( C_1 \) denote the underlying copula of a multivariate random vector \( \mathbf{X} \), and let \( \Pi(\mathbf{u}) = \prod_{j=1}^d u_j \) represent the product copula. For any \( \alpha \in \mathcal{A} \), the CMI of order \( \alpha \) is defined as 
\[
\mu_{\alpha}(C) = \Delta_{\alpha}(C \| \Pi) = \int_{\mathbb{I}^d} C(\mathbf{u}) \log_{[\alpha]}\left(\frac{C(\mathbf{u})}{\Pi(\mathbf{u})}\right) \, d\mathbf{u} - \frac{\rho_d(C)}{2^d c_d},
\]
where \( \rho_d(\cdot) \) is the multivariate Spearman's correlation. In the limiting case as \( \alpha \to 1 \), we have
\[
\mu(C) = \int_{\mathbb{I}^d} C(\mathbf{u}) \log\left(\frac{C(\mathbf{u})}{\Pi(\mathbf{u})}\right) \, d\mathbf{u} - \frac{\rho_d(C)}{2^d c_d}.
\]
This limiting case is referred to as cumulative mutual information. The proposed CMI provides an alternative to existing correlation measures. The existing correlation measures, such as Pearson’s correlation, are limited to linear relationships, while Spearman’s and Kendall’s correlations capture monotonic relationships but are primarily suited for bivariate cases. The proposed measure, on the other hand, quantifies deviations from independence to stronger dependence in any dimension, making it a robust candidate for dependency analysis in multivariate contexts. The application of CMI of order \( \alpha \) is illustrated in the subsequent section. 
We conclude this section by presenting a few examples of the proposed CCTD and CMI for well-known copulas.

 \begin{example}
     The CCTD measure between the FGM copula $C(u_1,u_2) = u_1u_2\left(1 + \theta (1-u_1)(1-u_2)\right)$ and Fr{\'e}chet-Hoeffding upper bound copula $M(u_1,u_2)=\min\{u_1,u_2\}$ is
    \begin{align*}
    \Delta_{\alpha}(C||M)=&\frac{2}{\alpha-1}\sum_{t=0}^{\infty}\binom{\alpha+t-1}{t}\theta^t\left[\frac{\beta(\alpha+1,t+1)}{(t+1)(t+2)}-\frac{\beta(\alpha+1,2t+1)}{(t+1)}+\frac{\beta(\alpha+1,2t+2)}{(t+2)(t+2)}\right]\\
    &\hspace{1cm}-\frac{\alpha(\theta+9)}{36(\alpha-1)}+\frac{1}{3}.
     \end{align*} 
 \end{example}
  \begin{example}
    The CCTD of the Gumbel-Barnett copula
     $$C(u_1,u_2)=u_1u_2\exp\{-\phi\log(u_1)\log(u_2)\}, \theta\in\mathbb{I},$$ then
     \begin{align*}
          \Delta_{\alpha}(C||\Pi)=\mu_{\alpha}(C)=&-\frac{e^{\frac{4}{(\alpha-1)\phi}}E_i\left(\frac{-4}{\phi(\alpha-1)}\right)}{\phi}+\frac{\alpha e^{\frac{4}{(\alpha-1)\phi}}E_i\left(\frac{-4}{\phi}\right)}{(\alpha-1)\phi}+\frac{1}{4}.
     \end{align*}
    We use the results of \cite{yela2018estimating} for computing the above intergals and $E_i(\cdot)$ is the well-known exponential integral function. 
    \end{example}
    \begin{example}
    The CCTD between the $d-$variate product copula and the $d$-variate Fr{\'e}chet-Hoeffding upper bound copula is 
    \begin{align*}
      \Delta_{\alpha}(\Pi||M)= \frac{d!}{2(\alpha-1)\prod_{j=1}^{d-1}(j(\alpha+1)+2)} -\frac{\alpha}{2^d(\alpha-1)}+\frac{1}{d+1}.
    \end{align*}
\end{example}
\begin{example}
   The CCTD of the Cuadras-Aug{\'e} copula $C(\mathbf{u}) = \prod_{i=1}^d u_{[i]}^{\gamma_i}$ is given by
   \begin{align*}
       \Delta_{\alpha}(C||\Pi)=\mu_{\alpha}(C)=\frac{d!}{(\alpha-1)}\left[\frac{1}{\prod_{i=1}^d\omega_1(i)}-\frac{\alpha}{\prod_{i=1}^d\omega_2(i)}\right]+\frac{1}{2^d},
   \end{align*}
   where $\omega_1(i)$ and $\omega_2(i)$ satisfies the recurrence relation given by
   \begin{align*}
       \omega_1(i)=&\omega_1(i-1)+(\theta_i-1)\alpha+1,\\
       \omega_2(i)=&\omega_2(i-1)+\theta_i+1,
   \end{align*}
   for $i=2,3,\dots,d$ with $\omega_1(1)=(\theta_1-1)\alpha+1$ and $\omega_2(1)=\theta_1+1$
\end{example}
\section{Application}\label{sec6}
Here, we explore the applications of the proposed mutual information measure in two different areas: testing for the mutual independence of continuous random variables and its relevance in the finance sector as an economic indicator.  

\subsection{Test for the Mutual Independence of Continuous Random Variables} \label{subsec1}
In multivariate data analysis, the assumption of mutual independence is frequently encountered. For such cases, Pearson’s correlation test is commonly used under the assumption of bivariate normality. Non-parametric tests, such as Spearman’s and Kendall’s correlation tests, are often used when the relationship between variables is monotonic. However, these tests are primarily designed to test the pairwise correlation for specific types of relationships and are often misused as tests for independence. 

\par Current research focused on empirical copula process-based tests for independence. The foundational idea was introduced by \cite{deheuvels1979fonction}. The Cram\'{e}r-von Mises and Kolmogorov-Smirnov functionals are widely used for testing mutual independence among random variables. For further details, we recommend \cite{deheuvels1979fonction}, \cite{genest2004test1}, \cite{genest2007asymptotic3}, \cite{kojadinovic2009tests4}, \cite{5belalia2017testing}, \cite{6herwartz2020nonparametric}, and \cite{7nasri2024tests}. 

Further, we propose using the CMI measure as a test statistic for testing the mutual independence among continuous random variables. We also compare the power of the proposed test with existing independence tests based on the Cram\'{e}r-von Mises and Kolmogorov-Smirnov statistics. To illustrate its practicality, we apply our test to a real dataset.

Let \(\mathbf{X} = (X_1, X_2, \dots, X_d)\) be a \(d\)-variate continuous random vector with an underlying copula \(C\). The copula \(C\) can be approximated by the empirical copula \(\hat{C}_n\), based on \(n\) random samples \(\mathbf{X}_1, \mathbf{X}_2, \dots, \mathbf{X}_n\), as defined in Eq.~(\ref{empirical}).  
To measure dependence, we consider the non-parametric cumulative mutual information (CMI). For mathematical simplicity, we take \(\alpha = 2\), yielding
\begin{align}
    \mu_2(\hat{C}_n) 
    =& \int_{\mathbb{I}^d} \hat{C}_n(\mathbf{u}) \log_{[2]}\left(\frac{\hat{C}_n(\mathbf{u})}{\Pi(\mathbf{u})}\right) \, d\mathbf{u} - \frac{\rho_d(\hat{C}_n)}{2^d c_d}\nonumber\\
    =& \int_{\mathbb{I}^d} \frac{\left(\hat{C}_n(\mathbf{u}) - \Pi(\mathbf{u})\right)^2}{\Pi(\mathbf{u})} \, d\mathbf{u}\nonumber\\
    =& \frac{1}{n^2} \sum_{i=1}^n \sum_{j=1}^n \prod_{k=1}^d \left[-\log\left(\max\left\{\frac{R_{ik}}{n+1}, \frac{R_{jk}}{n+1}\right\}\right)\right] - \frac{2}{n} \sum_{i=1}^n \prod_{k=1}^d \left[1 - \frac{R_{ik}}{n+1}\right] + \frac{1}{2^d}, \label{mu2}
\end{align}
where \(R_{ik}\) represents the rank of the \(k\)-th component of the \(i\)-th observation. 

Let \(\mathbf{X}_1, \mathbf{X}_2, \dots, \mathbf{X}_n\) be \(n\) random samples from a common multivariate population. We aim to test the null hypothesis \(H_0\) that the components of the multivariate population are mutually independent, i.e., the underlying copula is the product copula \(\Pi(\mathbf{u}) = \prod_{k=1}^d u_k\). Using the definition of non-parametric CMI from Eq.~(\ref{mu2}), we propose the following test statistic
\begin{align}
    \chi^2_{\text{div}}(n) 
    = n \mu_2(\hat{C}_n) 
    =& \frac{1}{n} \sum_{i=1}^n \sum_{j=1}^n \prod_{k=1}^d \left[-\log\left(\max\left\{\frac{R_{ik}}{n+1}, \frac{R_{jk}}{n+1}\right\}\right)\right]- 2 \sum_{i=1}^n \prod_{k=1}^d \left[1 - \frac{R_{ik}}{n+1}\right] + \frac{n}{2^d}. \label{stat}
\end{align}
Since we set \(\alpha = 2\), we call this the \emph{$\chi^2$ divergence test for mutual independence}, and denote the test statistic by \(\chi^2_{\text{div}}(n)\), where \(n\) is the sample size.

To study the asymptotic behavior of the proposed test under \(H_0\), the following lemma, discussed in \cite{fermanian2004weak}, \cite{tsukahara2005semiparametric}, and \cite{kojadinovic2009tests4} is useful.

\begin{lemma}\label{lem}
Let \(C\) be a \(d\)-dimensional copula. Let \(L_{\infty}(\mathbb{I}^d)\) denote the Banach space of real-valued bounded functions defined on \(\mathbb{I}^d\), equipped with the supremum norm. If \(C\) has continuous partial derivatives for every \(\mathbf{u} \in \mathbb{I}^d\), then the empirical process
\[
\mathbb{Z}_n(\mathbf{u}) = \sqrt{n}\left(\hat{C}_n(\mathbf{u}) - C(\mathbf{u})\right)
\]
converges weakly in \(L_{\infty}(\mathbb{I}^d)\) to the tight centered Gaussian process
\[
\mathbb{Z}(\mathbf{u}) = \Gamma(\mathbf{u}) - \sum_{i=1}^d \partial_i C(\mathbf{u}) \Gamma(\mathbf{u}_i),
\]
where \(\partial_i C(\mathbf{u})\) is the \(i\)-th partial derivative of \(C\), \(\mathbf{u}_i = (1, \dots, 1, u_i, 1, \dots, 1)\) with \(u_i\) in the \(i\)-th position, and \(\Gamma(\mathbf{u})\) is a tight centered Gaussian process on \(\mathbb{I}^d\) with covariance function
\[
\Sigma(\mathbf{u}, \mathbf{v}) = C(\mathbf{u} \wedge \mathbf{v}) - C(\mathbf{u})C(\mathbf{v}),
\]
where \(\mathbf{u} \wedge \mathbf{v} = \big(\min(u_1, v_1), \dots, \min(u_d, v_d)\big)\).
\end{lemma}

Using the Lemma \ref{lem} and the application of the continuous mapping theorem, we have the following theorem. 
\begin{theorem}
Let \(\mathbf{X}_1, \mathbf{X}_2, \dots, \mathbf{X}_n\) be \(n\) random samples from a multivariate population. Then, under the null hypothesis of mutual independence, the test statistic \(\chi^2_{\text{div}}(n)\) (as given in Eq.~(\ref{stat})) converges in distribution to 
\[
\int_{\mathbb{I}^d} \frac{\mathbb{Z}^2(\mathbf{u})}{\Pi(\mathbf{u})} \, d\mathbf{u},
\]
where 
\[
\mathbb{Z}(\mathbf{u}) = \Gamma\left(\mathbf{u}\right) - \sum_{i=1}^d \Gamma(\mathbf{u}_i) \prod_{\substack{j=1 \\ j \neq i}}^d \Pi(\mathbf{u}_j),
\]
is a tight centered Gaussian process with \(\mathbf{u}_i = (1, \dots, 1, u_i, 1, \dots, 1)\) represents the vector with the \(i\)-th component equal to \(u_i\) and all other components equal to 1 for \(i = 1, 2, \dots, d\). The process \(\Gamma(\mathbf{u})\) is a tight centered Gaussian process on \(\mathbb{I}^d\) with covariance function 
\[
\Sigma(\mathbf{u}, \mathbf{v}) = \mathbb{E}\left[\Gamma(\mathbf{u})\Gamma(\mathbf{v})\right] = \Pi(\mathbf{u} \wedge \mathbf{v}) - \Pi(\mathbf{u})\Pi\mathbf{v}),
\]
where \(\mathbf{u} \wedge \mathbf{v} = \left(\min(u_1, v_1), \dots, \min(u_d, v_d)\right)\).
\end{theorem}
Now, we will discuss the computation of p-values of the proposed test. Since the the distribution of the proposed test statistic $\chi^2_{\text{div}}(n)$ based on $n$ random samples is complex in nature, even in the asymptotic case, we employ the bootstrapping procedure to compute the approximate p-values.  The validity of the proposed approach is discussed in \cite{genest2008validity}. Let \(\mathbf{X}_1, \mathbf{X}_2, \dots, \mathbf{X}_n\) be \(n\) random samples from a multivariate population. Let $\mathbf{D}=\left(\mathbf{X}_1, \mathbf{X}_2, \dots, \mathbf{X}_n\right)'$ be the data matrix. Then the procedure for computing the p-values is discussed as follows.
\begin{enumerate}[Step 1:]
    \item Convert the data matrix \(\mathbf{D} = [\mathbf{X}_1, \mathbf{X}_2, \dots, \mathbf{X}_n]\) to the rank matrix \(\mathbf{R} = [R_{ik}]\), where \(R_{ik}\) is the rank of the \(k\)-th component of the \(i\)-th observation (i.e., $\mathbf{X}_i$). If ties occur, break them randomly.
    \item Calculate the test statistic \(\chi^2_{\text{div}}(n)\) using the formula
    \[
    \chi^2_{\text{div}}(n) = \frac{1}{n} \sum_{i=1}^n \sum_{j=1}^n \prod_{k=1}^d \left[-\log \left(\max \left\{ \frac{R_{ik}}{n+1}, \frac{R_{jk}}{n+1} \right\} \right)\right] 
    - 2 \sum_{i=1}^n \prod_{k=1}^d \left[1 - \frac{R_{ik}}{n+1}\right] + \frac{n}{2^d}.
    \]
    \item Generate \(B\) random samples of size \(n\) from the product copula. For each random sample, compute the test statistic \(\chi^2_{\text{div}}(n_b)\), \(b = 1, 2, \dots, B\).
    \item Arrange the computed bootstrap test statistics in ascending order
    \[
    \chi^2_{\text{div}}(n_{(1)}) \leq \chi^2_{\text{div}}(n_{(2)}) \leq \dots \leq \chi^2_{\text{div}}(n_{(B)}).
    \]
    \item Estimate the \(p\)-value associated with the observed test statistic \(\chi^2_{\text{div}}(n)\) as:
    \[
    \text{\(p\)-value} = \frac{1}{B} \sum_{b=1}^B \mathbf{1}\left\{\chi^2_{\text{div}}(n_{(b)}) \geq \chi^2_{\text{div}}(n)\right\},
    \]
    where \(\mathbf{1}\{\cdot\}\) is the indicator function.
\end{enumerate}
Now, we conduct the simulation study to evaluate the performance of the proposed model. We generate $10,000$ samples of various sizes and compute the power of the proposed test with the alternative copula such as Clayton, FGM, Frank, Normal and Student $t$ of various Kendall's Tau. We compare our results with Cram\'{e}r-von Mises (CVM) statistics which is given by
$$S_n=\frac{n}{3^d}-\frac{1}{2^{d-1}}\sum_{i=1}^n\prod_{k=1}^d\left(1-\left(\frac{R_{ik}}{n+1^2}\right)^2\right)+\frac{1}{n}\sum_{i=1}^n\sum_{j=1}^n\prod_{k=1}^d\left(1-\max\left\{\frac{R_{ik}}{n+1},\frac{R_{ik}}{n+1}\right\}\right)$$
and Kolmogorov-Smirnov (KS) statistics given by 
$$K_n=\sqrt{n}\underset{\mathbf{u}\in\mathbf{I}^d}{\sup} \ |\hat{C}_n(\mathbf{u})-C(\mathbf{u})|.$$
Note, that the explicit statistic of $K_n$ is often challenging and we approximate by its sample counterparts. The results are presented in Table \ref{tabl}.
\begin{landscape}
\begin{table}[ht]
    \centering
    \scalebox{0.75}{
    \begin{tabular}{|c|c|ccc|ccc|ccc|ccc|ccc|}\hline
        \multirow{2}{*}{True Copula} & \multirow{2}{*}{Test} 
        & \multicolumn{3}{c|}{\(\tau=-0.2\)} & \multicolumn{3}{c|}{\(\tau=-0.1\)} 
        & \multicolumn{3}{c|}{\(\tau=0\)} & \multicolumn{3}{c|}{\(\tau=0.1\)} 
        & \multicolumn{3}{c|}{\(\tau=0.2\)} \\
        & & \(n=50\) & \(n=100\) & \(n=150\) & \(n=50\) & \(n=100\) & \(n=150\) 
        & \(n=50\) & \(n=100\) & \(n=150\) & \(n=50\) & \(n=100\) & \(n=150\) 
        & \(n=50\) & \(n=100\) & \(n=150\) \\\hline
        
        \multirow{3}{*}{Clayton} 
        & CVM Test       & 0.5394 & 0.8427 & 0.8603 & 0.1694 & 0.2984 & 0.4341 & 0.037 & 0.049 & 0.0518 & 0.1476 & 0.2776 & 0.4098 & 0.4984 & 0.8104 & 0.9386 \\
        & KS Test        & 0.1143 & 0.4531 & 0.4779 & 0.0267 & 0.0860 & 0.1663 & 0.0494 & 0.0487 & 0.0477 & 0.1743 & 0.2813 & 0.3779 & 0.4892 & 0.4926 & 0.8991 \\
        & Proposed Test  & 0.7065 & 0.9624 & 0.9692 & 0.2184 & 0.4250 & 0.5923 & 0.0580 & 0.0513 & 0.0492 & 0.2142 & 0.3952 & 0.5572 & 0.6491 & 0.8148 & 0.9838 \\\hline
        
        \multirow{3}{*}{FGM} 
        & CVM Test       & 0.5443 & 0.8344 & 0.8483 & 0.1646 & 0.2901 & 0.4383 & 0.0537 & 0.0471 & 0.0502 & 0.1646 & 0.3033 & 0.4387 & 0.5248 & 0.8501 & 0.9542 \\
        & KS Test        & 0.1387 & 0.7562 & 0.5187 & 0.0298 & 0.0947 & 0.1969 & 0.0511 & 0.0516 & 0.0489 & 0.0298 & 0.3155 & 0.4282 & 0.5263 & 0.8080 & 0.9196 \\
        & Proposed Test  & 0.5336 & 0.9187 & 0.8290 & 0.1669 & 0.2820 & 0.3958 & 0.0585 & 0.0486 & 0.0513 & 0.1292 & 0.2478 & 0.3523 & 0.4313 & 0.7740 & 0.9151 \\\hline
        
        \multirow{3}{*}{Frank} 
        & CVM Test       & 0.5220 & 0.8267 & 0.8386 & 0.1675 & 0.2978 & 0.4336 & 0.0499 & 0.0512 & 0.0512 & 0.1646 & 0.3038 & 0.4375 & 0.5170 & 0.8350 & 0.9489 \\
        & KS Test        & 0.1359 & 0.8080 & 0.5107 & 0.0270 & 0.0984 & 0.1955 & 0.0514 & 0.0590 & 0.0483 & 0.2015 & 0.3217 & 0.4281 & 0.5231 & 0.7977 & 0.9223 \\
        & Proposed Test  & 0.5078 & 0.7740 & 0.8092 & 0.1673 & 0.2973 & 0.3893 & 0.0545 & 0.0612 & 0.0499 & 0.1347 & 0.2571 & 0.3596 & 0.4375 & 0.7667 & 0.9178 \\\hline
        
        \multirow{3}{*}{Normal} 
        & CVM Test       & 0.5086 & 0.8052 & 0.8187 & 0.1570 & 0.2731 & 0.4037 & 0.0621 & 0.0492 & 0.0511 & 0.1586 & 0.2824 & 0.4066 & 0.4897 & 0.8012 & 0.9312 \\
        & KS Test        & 0.1083 & 0.4287 & 0.4258 & 0.0234 & 0.0790 & 0.1586 & 0.0624 & 0.0481 & 0.0486 & 0.1858 & 0.2884 & 0.3790 & 0.4727 & 0.7321 & 0.8739 \\
        & Proposed Test  & 0.5105 & 0.7985 & 0.8111 & 0.1649 & 0.2799 & 0.3814 & 0.0550 & 0.0465 & 0.0541 & 0.1311 & 0.2556 & 0.3604 & 0.4434 & 0.7638 & 0.9106 \\\hline
        
        \multirow{3}{*}{Student's \(t\)} 
        & CVM Test       & 0.4987 & 0.7885 & 0.8036 & 0.1805 & 0.2937 & 0.4305 & 0.0580 & 0.0521 & 0.0478 & 0.1772 & 0.3050 & 0.4260 & 0.4958 & 0.7927 & 0.9261 \\
        & KS Test        & 0.1126 & 0.4318 & 0.4378 & 0.0292 & 0.0863 & 0.1743 & 0.0610 & 0.0528 & 0.0498 & 0.2093 & 0.3063 & 0.3926 & 0.4926 & 0.7600 & 0.8825 \\
        & Proposed Test  & 0.4960 & 0.7638 & 0.7878 & 0.1930 & 0.3026 & 0.4087 & 0.0590 & 0.0492 & 0.0501 & 0.1909 & 0.3527 & 0.4785 & 0.4976 & 0.7947 & 0.9279 \\\hline
    \end{tabular}}
    \caption{Power comparison of tests for different true copulas, Kendall's \(\tau\), and sample sizes \(n\).}
    \label{tabl}
\end{table}
\end{landscape}

From Table \ref{tabl}, it is clear that the proposed test rejection power is superior when the alternative copula is Clayton compared to other tests. The results show a significant improvement over CVM and KS tests. The proposed test also performs better for the Student's \(t\) copula across various Kendall's \(\tau\). For the remaining copulas, the power of the proposed test is comparable, making it a strong candidate for testing the mutual independence of random variables. 

\begin{figure}[ht]
    \centering
    \includegraphics[width=0.5\linewidth]{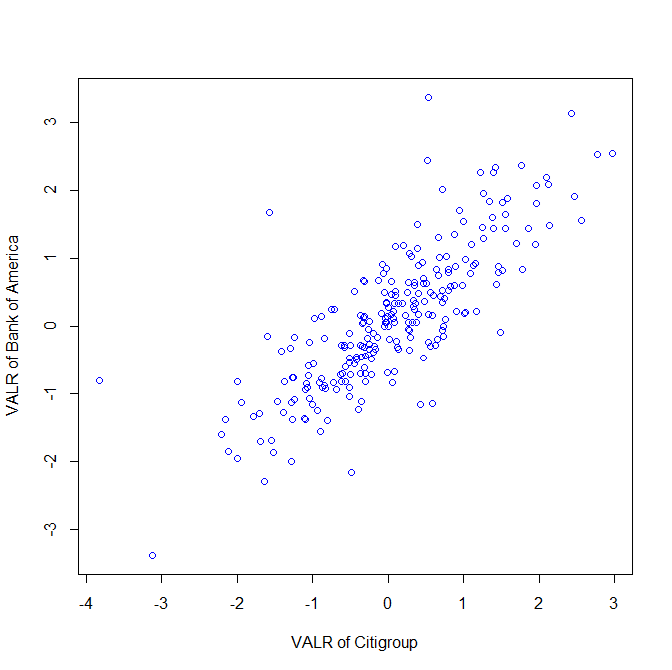}
    \caption{Scatterplot of volatility-adjusted log returns of Citigroup and Bank of America.}
    \label{valr}
\end{figure}

We now apply our test to a real dataset. This dataset comprises 249 observations of the volatility-adjusted log returns (VALR) of two banks, Citigroup and Bank of America, for the year 2012. The \textbf{Banks} dataset is freely accessible in the R software within the \texttt{gofCopula} package. The scatterplot of the data is presented in Figure~\ref{valr}. From Figure~\ref{valr}, it is evident that there is a strong positive dependence between the VALR of the two banks. The proposed test supports this observation, yielding a test statistic of \(12.089\) and a \(p\)-value approximately equal to zero.

\subsection{Application in Financial Time Series}\label{subsec2}
\begin{figure}[ht]
    \centering
    \includegraphics[width=0.6\linewidth]{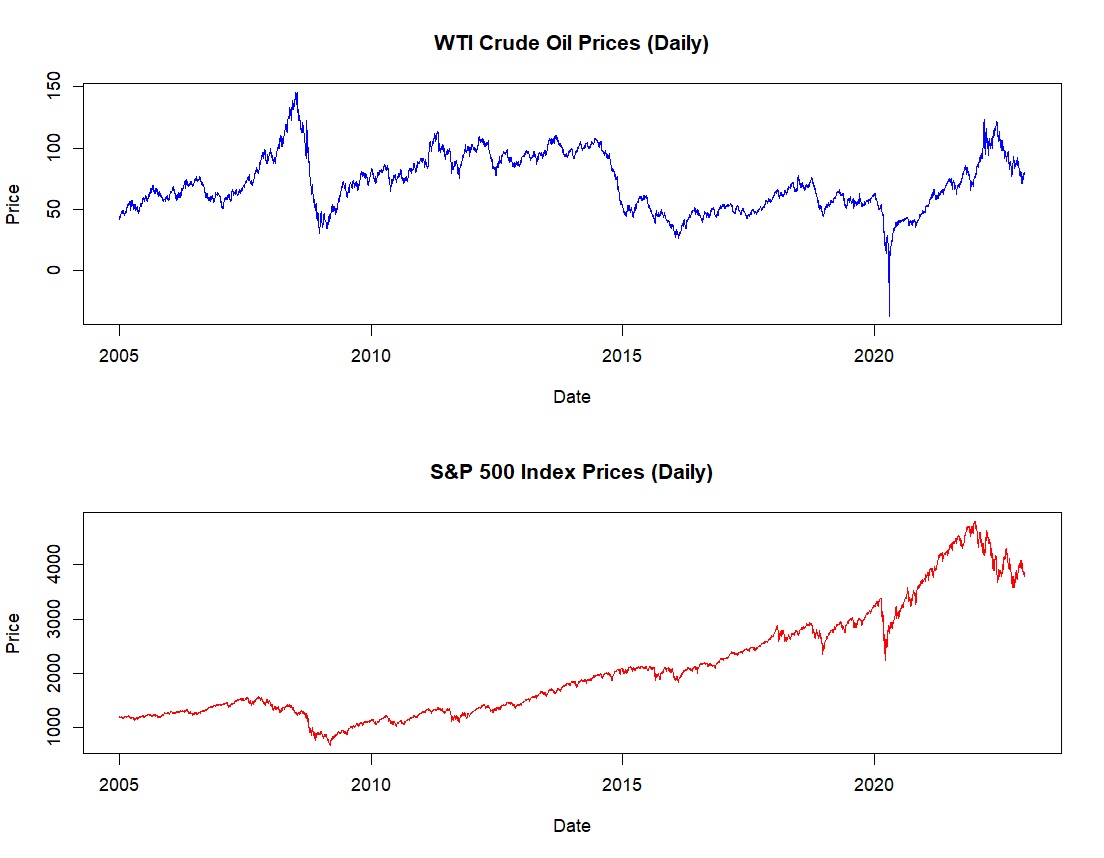}
    \caption{Daily data of Crude Oil and S\&P 500 index.}
    \label{daily}
\end{figure}

\begin{figure}[ht]
    \centering
    \includegraphics[width=0.6\linewidth]{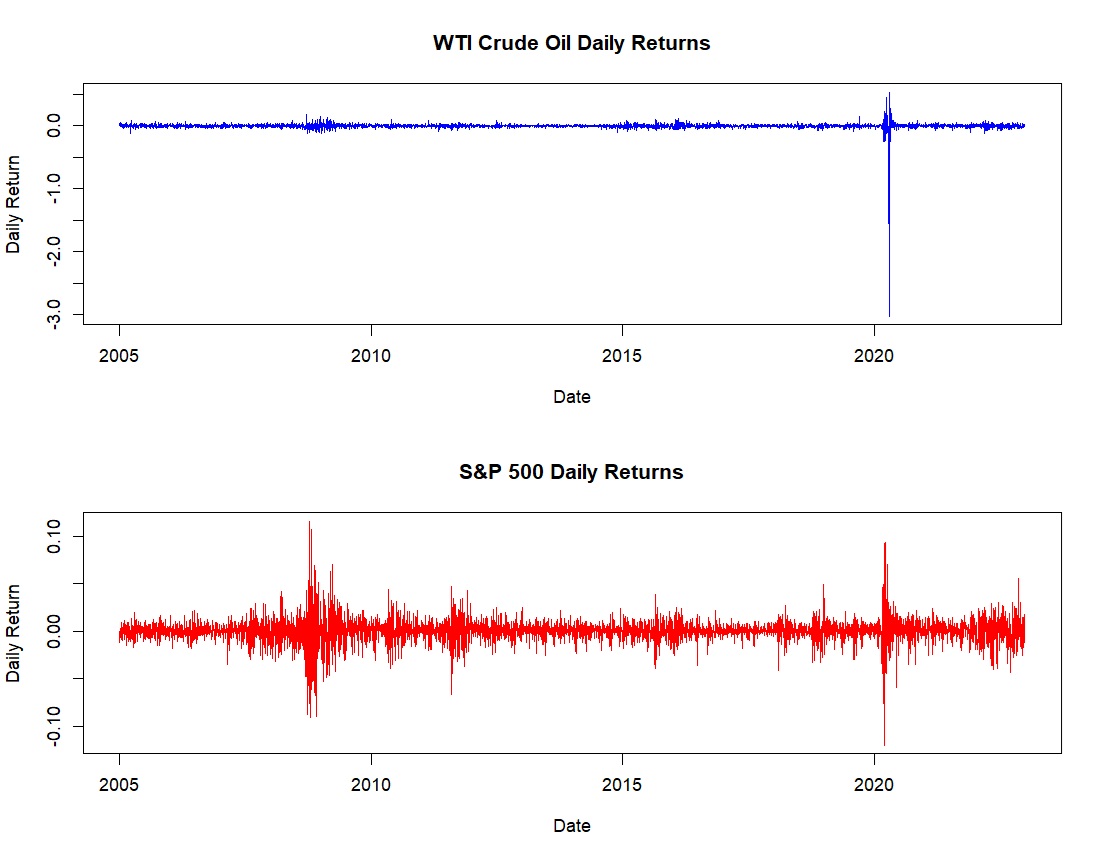}
    \caption{Daily returns of Crude Oil and S\&P 500 index.}
    \label{dailyr}
\end{figure}

In this subsection, we present our proposed CMI as an economic indicator. We consider the daily price returns of Crude Oil and the S\&P 500 index during the period from January 2, 2005, to December 31, 2022. The plots of daily data and daily returns are shown in Figures~\ref{daily} and \ref{dailyr}, respectively.

\begin{figure}[h!]
    \centering
    \includegraphics[width=1\linewidth]{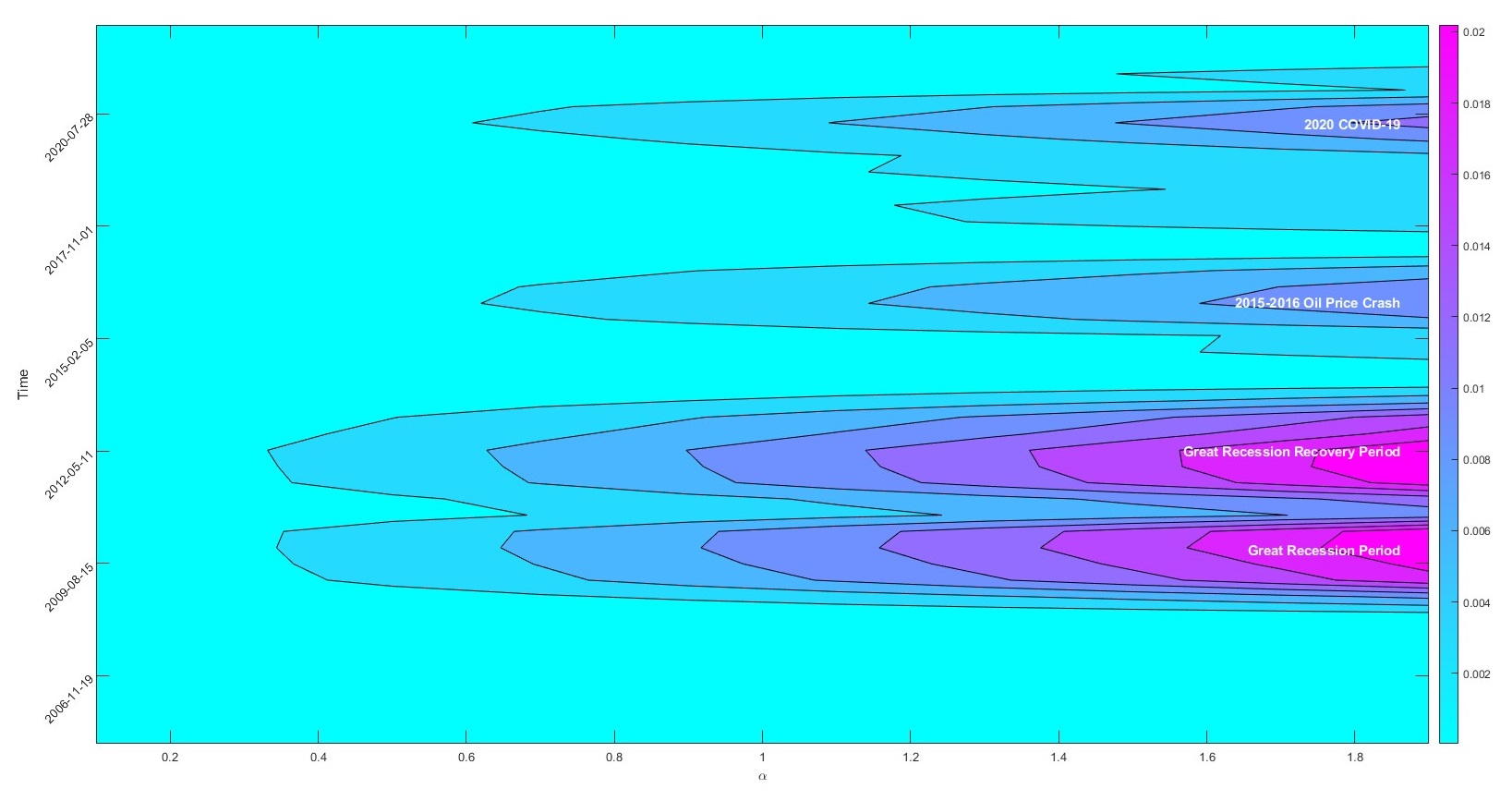}
    \caption{Contour plot of the proposed CMI for different values of $\alpha$.}
    \label{cctent}
\end{figure}

The data for Crude Oil was obtained from the Federal Reserve Economic Data (FRED), and the data for the S\&P 500 index was sourced from Yahoo Finance. We compute the proposed CMI using an overlapping sliding time window of 200 data points, with a shift size of 100 points. This approach reveals the evolution of the series over time and identifies any mutual information between the daily price returns of WTI Crude Oil and the S\&P 500 index. The contour plot of the proposed CMI for different values of $\alpha$ is shown in Figure~\ref{cctent}.

During the period from 2005 to 2022, several significant financial events occurred:
\begin{itemize}
    \item The global economic recession from 2008 to 2009 and slow recovery impacted crude oil prices from 2010 to 2012.
    \item The oil price crash occurred between 2015 and 2016.
    \item The COVID-19 pandemic caused significant financial disruptions in 2020.
\end{itemize}

For more details, we refer to \cite{lyu2021time}, \cite{stocker2018benefit}, and news articles during these financial crisis periods. The proposed CMI measure effectively captures the financial crises, as evident from the results. For higher values of $\alpha$, the CMI increases, further supporting its potential as an economic indicator for financial crises.

\section{Conclusion and Future Directions}\label{sec7}

In this paper, we introduced a new non-additive dependence entropy called \textit{cumulative copula Tsallis entropy}. We discussed its mathematical properties, including bounds, copula ordering, and uniform convergence results. Additionally, we demonstrated that the proposed entropy generalizes the cumulative copula entropy introduced by \cite{arshad2024multivariateinformationmeasurescopulabased}. Using the empirical copula, we proposed a non-parametric estimator for the entropy and established its theoretical convergence as well as its convergence through Monte Carlo simulations.

To validate the utility of the proposed entropy in quantifying uncertainty in dependence structures, we examined Rulkov maps. Our findings indicate that the proposed entropy increases with periodicity and reaches its maximum in chaotic cases. To address the uncertainty arising from incorrect copula assumptions, we proposed a copula-based Kerridge inaccuracy measure, studied its properties (including triangular inequalities), and demonstrated its generalization of the results presented in \cite{hosseini2021discussion}. These concepts were illustrated using well-known copulas.

Furthermore, we introduced \textit{cumulative copula divergence} using Tsallis divergence. Based on this, a new mutual information measure, termed \textit{cumulative mutual information}, was proposed by leveraging its relationship with Kullback-Leibler divergence. This approach overcomes limitations in the existing copula density-based mutual information. The utility of this mutual information measure was demonstrated in two important statistical applications:
\begin{itemize}
    \item \textbf{Hypothesis testing}, specifically for mutual independence among random variables.
    \item \textbf{Finance}, as an economic indicator for multivariate time series, providing a robust alternative to traditional correlation measures.
\end{itemize}
While our study focused on Tsallis entropy, the proposed methodology can be extended to other entropies, such as R\'{e}nyi entropy. Moreover, recent advancements using the M\"{o}bius decomposition of the empirical copula process have been shown to improve the power of tests based on the Cram\'{e}r-von Mises statistic. For further details, see \cite{deheuvels1979fonction}, \cite{genest2004test}, and \cite{kojadinovic2009tests4}. Incorporating similar techniques into our proposed test could significantly enhance its power, making it an interesting direction for future research.

\section*{Acknowledgment}

Mohd. Arshad would like to express his gratitude to the Science and Engineering Research Board (SERB), India
for financial support under the Core Research Grant CRG/2023/001230. Swaroop Georgy Zachariah is thankful to
the IIT Indore for providing financial support. Ashok Kumar Pathak would like to express his gratitude to the SERB,
India for financial support under the MATRICS research grant MTR/2022/000796. The authors are grateful to Ms. Rajni and Mr. Debdeep Roy for generously sharing the MATLAB codes used for generating Rulkov maps. Finally, the authors would like to thank Mr. Navanit A V for his valuable suggestions in the time series analysis.

\section*{Conflicts of interests}
The authors declare no potential conflict of interest


\bibliography{references}







\end{document}